\documentclass[12pt,preprint]{aastex}

\def\deg{\ifmmode^\circ\else$^\circ$\fi}

\def\mic{~$\mu$m}

\def\mic{$\mu${\rm m}}
\def\lir{{\rm L}$_{IR}$}

\def\arcs{\ifmmode {''}\else $''$\fi}
\def\arcm{\ifmmode {'}\else $'$\fi}
\def\parcs{\sa=.07em \sb=.03em
     \ifmmode $\rlap{.}$^{\scriptscriptstyle\prime\kern -\sb\prime}$\kern -\sa$
     \else \rlap{.}$^{\scriptscriptstyle\prime\kern -\sb\prime}$\kern -\sa\fi}
\def\parcm{\sa=.08em \sb=.03em
     \ifmmode $\rlap{.}\kern\sa$^{\scriptscriptstyle\prime}$\kern-\sb$
     \else \rlap{.}\kern\sa$^{\scriptscriptstyle\prime}$\kern-\sb\fi}

\def\Msun{M$_{\odot}$}

\def\Myr{\Msun/yr}

\def\lya{{\rm Ly}$\alpha$}

\def\spose#1{\hbox to 0pt{#1\hss}}
\def\simlt{\mathrel{\spose{\lower 3pt\hbox{$\mathchar"218$}}
     \raise 2.0pt\hbox{$\mathchar"13C$}}}
\def\simgt{\mathrel{\spose{\lower 3pt\hbox{$\mathchar"218$}}
     \raise 2.0pt\hbox{$\mathchar"13E$}}}
\def\lsim{\rlap{$<$}{\lower 1.0ex\hbox{$\sim$}}}
\def\gsim{\rlap{$>$}{\lower 1.0ex\hbox{$\sim$}}}

\begin{document}

\title{Far-Ultraviolet Imaging of the Hubble Deep Field North$^1$: Star Formation
in Normal Galaxies at $z<1$}

\altaffiltext{1}{Based on observations made with the NASA/ESA {\it Hubble Space Telescope}, obtained
from the Space Telescope Science Institute, which is operated by the 
Association of Universities for Research in Astronomy, Inc., under NASA
contract NAS 5-26555. These observations are associated with proposals 7410 and 9478. }

\author{H. I. Teplitz\altaffilmark{2}, 
B. Siana\altaffilmark{2}, 
T. M. Brown\altaffilmark{3},
R. Chary\altaffilmark{2},
J. W. Colbert\altaffilmark{2},
C. Conselice\altaffilmark{4},
D. F. de Mello\altaffilmark{5,6,7},
M. Dickinson\altaffilmark{8},
H. C. Ferguson\altaffilmark{3},
Jonathan P. Gardner\altaffilmark{5}
F. Menanteau\altaffilmark{7}
}

\altaffiltext{2}{Spitzer Science Center, MS 220-6, Caltech, Pasadena, CA 91125.  hit@ipac.caltech.edu}
\altaffiltext{3}{Space Telescope Science Institute, 3700 San Martin Drive, Baltimore, MD 21218}
\altaffiltext{4}{University of Nottingham, Nottingham, NG7 2RD, UK}
\altaffiltext{5}{Exploration of the Universe Division, Observational Cosmology Laboratory, Code 665, 
Goddard Space Flight Center, Code 681, Greenbelt, MD 20771}
\altaffiltext{6}{Department of Physics, Catholic University of America, 620 Michigan Avenue, Washington, DC 20064}
\altaffiltext{7}{Department of Physics and Astronomy, Johns Hopkins University, 3400 North Charles Street, Baltimore, MD 21218}
\altaffiltext{8}{NOAO, 950 N. Cherry Ave., Tucson AZ 85719}

\begin{abstract}
  
  We present far-ultraviolet (FUV) imaging of the Hubble Deep Field
  North (HDF-N) taken with the Solar Blind Channel of the Advanced
  Camera for Surveys (ACS/SBC) and the FUV MAMA detector of the Space
  Telescope Imaging Spectrograph (STIS) onboard the {\it Hubble Space
    Telescope}.  The full WFPC2 deep field has been observed at 1600
  Angstroms.  We detect 134 galaxies and one star down to a limit of
  $FUV_{AB}\sim 29$.  All sources have counterparts in the WFPC2
  image.  Redshifts (spectroscopic or photometric) for the detected
  sources are in the range $0<z<1$.  We find that the FUV galaxy
  number counts are higher than those reported by GALEX, which we
  attribute at least in part to cosmic variance in the small HDF-N
  field of view.  Six of the 13 {\it Chandra}\ sources at $z<0.85$\ in
  the HDF-N are detected in the FUV, and those are consistent with
  starbursts rather than AGN.  Cross-correlating with {\it Spitzer}\ 
  sources in the field, we find that the FUV detections show general
  agreement with the expected $L_{\rm{IR}}/L_{\rm{UV}}$\ vs.\ $\beta$\ 
  relationship.  We infer star formation rates (SFRs), corrected for
  extinction using the UV slope, and find a median value of 0.3 \Myr\ 
  for FUV-detected galaxies, with 75\%\ of detected sources have
  SFR$<1$\ \Myr.  Examining the morphological distribution of sources,
  we find that about half of all FUV-detected sources are identied as
  spiral galaxies.  Half of morphologically-selected spheroids at $z<
  0.85$\ are detected in the FUV, suggesting that such sources have
  significant ongoing star-formation in the epoch since $z\sim 1$.

\end{abstract}


\section{Introduction}

The star formation rate density of the Universe, integrated over all
galaxy populations, shows a sharp decline since redshifts near unity
\citep[e.g., ][]{Madau 1996, Madau 1998}.  While the precise shape of
the decline with redshift is still uncertain \citep{Lilly 1996, Hogg
  1998, Flores 1999, Wilson 2002}, its existence points to a
``downsizing'' in galaxies that host most of the star formation at
$z<1$\ \citep{Cowie 1996}.  The characteristics (morphology, mass,
luminosity) of these low redshift starbursts may explain the global
decline in star formation.  \cite{Wolf 2005}\ suggest that the decline
is dominated by decreasing star formation in normal spiral galaxies
rather than, for example, the decreasing rate of major mergers.

Ultraviolet (UV) emission is an indication of recent star formation in
a galaxy.  Despite absorption by dust, the rest-frame UV is strong
enough in the majority of star-forming galaxies to be detected in
current surveys \citep{Adelberger and Steidel 2000}.  UV detection can
distinguish star-forming from quiescent systems, and indicates the
amount of recent star formation, subject to the effects of extinction
by dust \citep[e.g.,][]{Calzetti 1994, Fitzpatrick 1986, Meurer 1999,
  Buat 2005}.  Far-ultraviolet (FUV) surveys, in particular, can
provide direct evidence of recently formed, massive stars in the
dominant populations at $z<1$\ \citep{Schiminovich 2005}.

Of particular interest is the the star-formation activity present in
early type galaxies.  Recent studies have shown that some galaxies
which appear morphologically to be spheroids have, nonetheless,
substantial ongoing star formation.  \cite{Yi 2005} find that at least
15\%\ of bright, local ellipticals show evidence of recent star
formation, ruling out pure monolithic collapse histories for at least
that fraction of such sources.  Similarly, studies of the internal
color variations in elliptical galaxies have shown almost one third of
them show gradients inconsistent with passive evolution \citep{Abraham
  1999, Menanteau 2001, Papovich 2003}.  The formation of spheroid
galaxies, then, is a crucial component of successful hierarchical
models.  \cite{Conselice 2005}\ find that the massive galaxies at
$z<1$, both spirals and ellipticals, likely have major-merger
progenitors at higher redshifts.  Nonetheless, a significant fraction
of stellar mass must still have formed since $z\sim 1$\ \citep{Bell
  2004, Dickinson 2003}.  The assembly of that additional stellar mass
should be detectable in UV surveys.

We present a far-ultraviolet (1600 \AA; FUV) imaging survey of the
Hubble Deep Field North \citep[HDF-N;][]{Williams 1996}.  The data
were taken in two surveys.  The first utilized the FUV camera of the
Space Telescope Imaging Spectrograph \citep[STIS;][]{Kimble 1998,
  Woodgate 1998}\ to survey a small part of the field.  That portion
of the data set has been been used to measure galaxy number-magnitude
counts \citep{Gardner 2000 GBF}\ and the
diffuse FUV background emission \citep{Brown 2000}.  We surveyed the
remaining area with the Solar Blind Channel (SBC) of the Advanced
Camera for Surveys \citep{Ford 1998}.  We outline the survey and data
reduction in Section \ref{sec: obs}\ and present the catalog in
Section \ref{sec: results}.  In Section \ref{sec: discussion}, we
discuss the properies of FUV-detected sources at other wavelengths,
the inferred star formation rates, and the implications of the
detection of elliptical galaxies.  Throughout, we assume a
$\Lambda$-dominated flat universe, with $H_0=70$\ km s$^{-1}$\ 
Mpc$^{-1}$, $\Omega_{\Lambda}=0.7, \Omega_{m}=0.3$.  Photometry is
presented with magnitudes on the AB system which is defined by $AB =
-2.5 log F_{\nu} - 48.6$, where $F_{\nu}$\ is given in units of ergs
cm$^{-2}$\ s$^{-1}$\ Hz$^{-1}$\ \citep{Oke 1971}.

\section{Observations and Data Reduction}
\label{sec: obs}

FUV imaging of the HDF-N was obtained in two HST General Observer
programs (No. 7410 with STIS and No. 9478 with ACS/SBC).  The ACS
survey is composed of fourteen 2-orbit pointings covering 3.77 square
arcminutes.  Each pointing consisted of 16$\times$640 seconds
exposures with dithers of $\sim$10 pixels.  The field of view of the
ACS/SBC detector is 34.6\arcs $\times 30.8$\arcs.  The images were
obtained with the long pass quartz filter (F150LP) with an effective
wavelength of 1614$\AA$ and $FWHM=177$\ \AA.  The STIS survey covered
1.02 square arcminutes in six pointings for a combined exposure time
of 124,330 seconds.  The field of view of the STIS detector is
25\arcs$\times 25$\arcs.  The images were obtained through the crystal
quartz filter (F25QTZ) with a central wavelength of 1595 \AA\ and
$FWHM=193$\ \AA.

\begin{figure}[t*]
\plotone{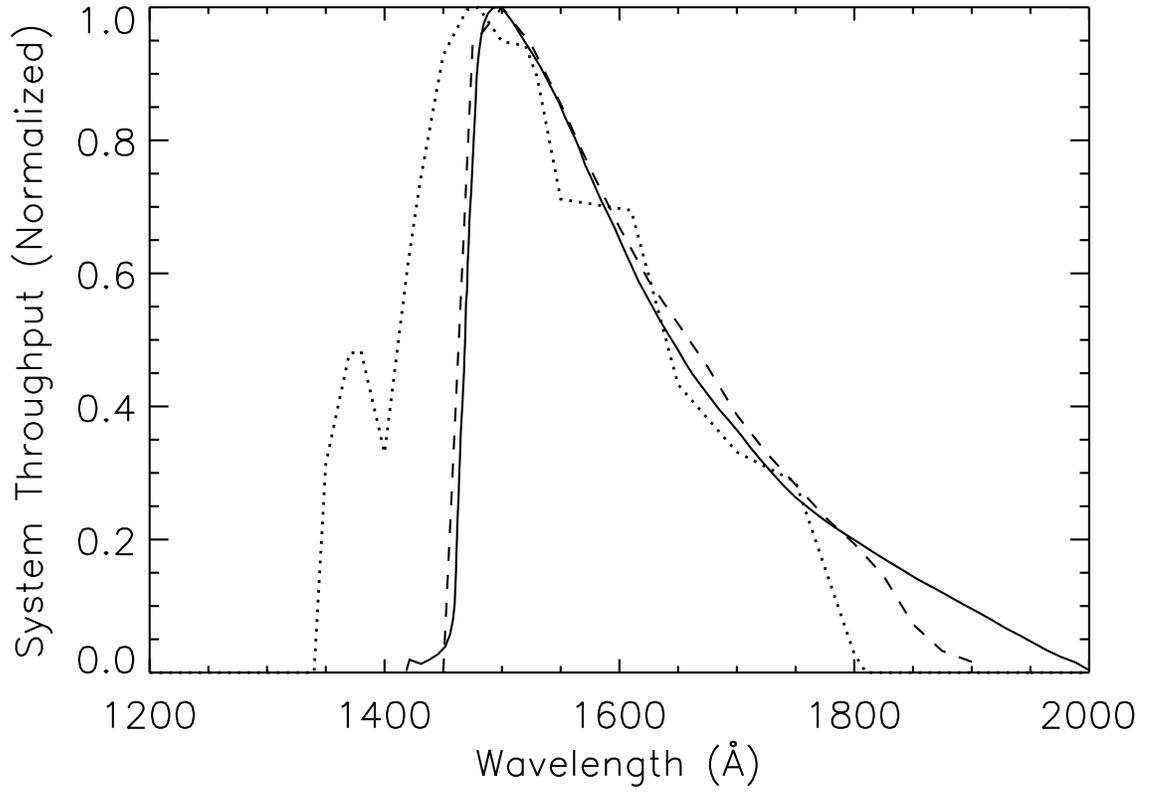}
\caption{ \label{fig: filter_curve} Total system throughput (filter+detector) for ACS/SBC F150LP (solid), 
STIS/FUVMAMA F25QTZ (dashed), and GALEX-FUV (dotted).}
\end{figure}

Both the STIS and ACS/SBC filters are long-pass filters with a short
wavelength cutoff at $\lambda < 1480$\ \AA.  The STIS and ACS/SBC
detectors are both Multi-Anode Microchannel Arrays (MAMAs) and have
similar spectral response curves which fall off slowly from 1500 to
1800 \AA.  Therefore, the combined filter+detector system response
curves of the STIS and ACS/SBC FUV configurations is similar, with the
only significant difference being that the ACS/SBC throughput is
non-zero at 1850 \AA\ $< \lambda <$\  2000 \AA\ (Figure \ref{fig: filter_curve}).

Data for both surveys were reduced following the procedure outlined in
\citet{Gardner 2000}.  Full details of the STIS data reduction are
given in that paper, and ACS-specific reductions are discussed here.

The MAMA detector has no read noise and is insensitive to cosmic rays.
The primary source of noise is dark current, which has two components.
When the temperature of the MAMA is below 20C, the dark current is
fairly uniform with an average count rate of $\sim 8\times 10^{-6}$
counts s$^{-1}$ pixel$^{-1}$.  However, as the SBC is being used, the
MAMA warms up and produces an additional temperature-dependent dark
``glow'' near the center of the detector.  Therefore, we subtract the
dark current in two stages.  First we subtract the primary calibration
dark which was made from darks collected at T$<$20C.  We then make a
residual dark by summing up all of the initial dark subtracted frames
and fitting a two dimensional, fifth-order spline curve.  The
isophotal segmentation maps from the HDF-North Wide-Field Planetary
Camera 2 (WFPC2) V+I images (Williams et al. 1996) were used to mask
areas contaminated by known objects using the {\it blot} capability of the
DRIZZLE package in IRAF.  We then subtracted the residual dark after
scaling to the average of a region near the peak of the dark ``glow''.
We find that $<2\%$ of the secondary dark remains after this second
subtraction.

\begin{figure}[t*]
\plotone{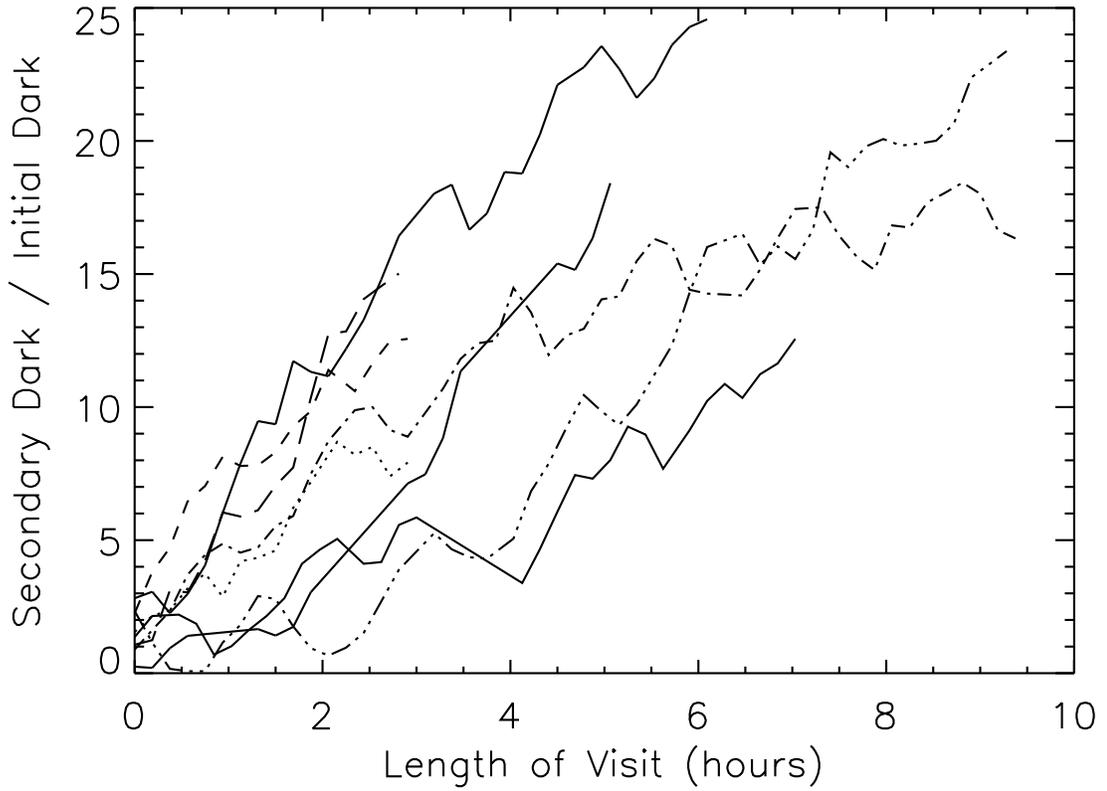}
\caption{ \label{fig: dark_rate} Count rate of the dark ``glow'' scaled to the dark rate at 
  $T<20C$ ($\sim 8\times 10^{-6}$ counts s$^{-1}$ pixel$^{-1}$) vs.
  time elapsed since the beginning of the visit.  Each line denotes a
  different visit to the field.  The count rate increases linearly
  with time at $\sim 2.5\times 10^{-5}$ counts$^{-1}$ s$^{-1}$
  pixel$^{-1}$ hour$^{-1}$.}
\end{figure}

The dark rate of the secondary ``glow'' near the center of the chip is
typically larger than that of the initial (i.e., at $T<20C$) dark rate
and is a function of the temperature of the MAMA tube.  The MAMA tube
gets warmer as the SBC is used, and we find that the count rate
increases linearly with time with the rate near the center of the chip
increasing by $\sim 2.5\times 10^{-5}$ counts$^{-1}$ s$^{-1}$
pixel$^{-1}$ hour$^{-1}$ (see Figure \ref{fig: dark_rate}).
Equivalently, this amounts to increasing the dark rate by an amount
equal to the ``cold'' dark rate every 20 minutes.  Some of our
scheduled visits were longer than 6 hours with entire pointings done
at the end of the visit.  Therefore the dark rate for these pointings
was a factor of 10-20 times larger than in those taken at the
beginning of a visit, resulting in a decrease in sensitivity of $\sim
$1-2 magnitudes in the ``glow'' regions.  Therefore, large SBC programs
in the future would greatly benefit from segmenting their observations into 
multiple visits.

Standard calibration files were used for flat fielding, geometric
distortion correction and photometric calibration.
Individual reduced images were registered and summed using the DRIZZLE
package in IRAF\footnote{IRAF is distributed by NOAO, which is
  operated by AURA Inc., under contract to the NSF}.  Both ACS and
STIS data were drizzled to the pixel scale of the HDF-N WFPC2 data
products (0.03985$''$ pixel$^{-1}$), and matched to the WFPC2 pixel
positions.  Shifts between the 16 dithered positions of each pointing
were assumed to be exactly as commanded.  This assumption is
reasonable, given the small-offset accuracy of HST (3-5 mas).  The
registration of the 14 pointings was done by matching the FUV sources
to the $B_{450}$ WFPC2 image and computing shifts, rotation, and
scaling with the GEOMAP routine in IRAF.  For all 14 pointings, the
rms fit was $<0.03''$.

As discussed by \cite{Gardner 2000}\ and \cite{Brown 2000}, the dark current is the
principle source of noise in MAMA imaging.  The individual frames were
weighted by the square of the exposure time, divided by the total dark
(initial + secondary).  As the dark count scales with exposure time,
these weight maps scale linearly with the ratio of exposure time to
dark rate.  Therefore, the final weight maps are the square of the
signal-to-noise ratio for objects fainter than the background.

Photometry was performed by summing the pixel values within the source
areas defined by the 3.25$\sigma$ isophotes (where $\sigma$ is the rms
background noise) in the V+I WFPC2 segmentation map produced by
SExtractor \citep{Bertin and Arnouts 1996}.  The V+I image was used
because it is the most sensitive image, and therefore the 3.25$\sigma$
isophotes are contiguous (i.e. individual galaxies are not broken into
several isophotes) and encompass the large majority of the galactic
light for all but the faintest galaxies.  To validate this method we
extracted fluxes in the $B_{450}$ image with 3.25$\sigma$ and
0.65$\sigma$ V+I isophotes.  The fluxes derived with the smaller
3.25$\sigma$\ isophotes missed 10--15\%\ of the flux within the larger
0.65$\sigma$\ isophote, but were significantly less noisy.  We
therefore use the smaller isophotes for detection and the ratio of the
two fluxes (in the F450W image) as an aperture correction.  Again, we
verified that these ``aperture corrected'' UV fluxes agreed well with
the fluxes derived in the larger aperture, but with smaller errors.
For the few objects with aperture corrections larger than 40\%, the
larger apertures were used.

\section{Results}
\label{sec: results}

Figure \ref{fig: FUV-mosaic} shows the fully reduced, registered FUV
mosaic.  We detect 128 sources above a signal-to-noise ratio,
SNR$>3.5$. We add an additional 7 sources by hand because their UV
flux is more compact and the larger V+I segmentation map causes large
errors in the flux estimates.  The FUV properties of the 135 sources
are given in Table \ref{tbl-1}.  The detection limits vary
significantly between pointings due to large variations in dark
``glow''.  In those regions least affected by the dark ``glow'', the
3.5$\sigma$ limiting magnitudes are $FUV_{AB} = 29.2$ in the STIS
survey and $FUV_{AB} = 28.8$ in the ACS Survey for a 1$''$ diameter
aperture.

\begin{figure}[t*]
\plotone{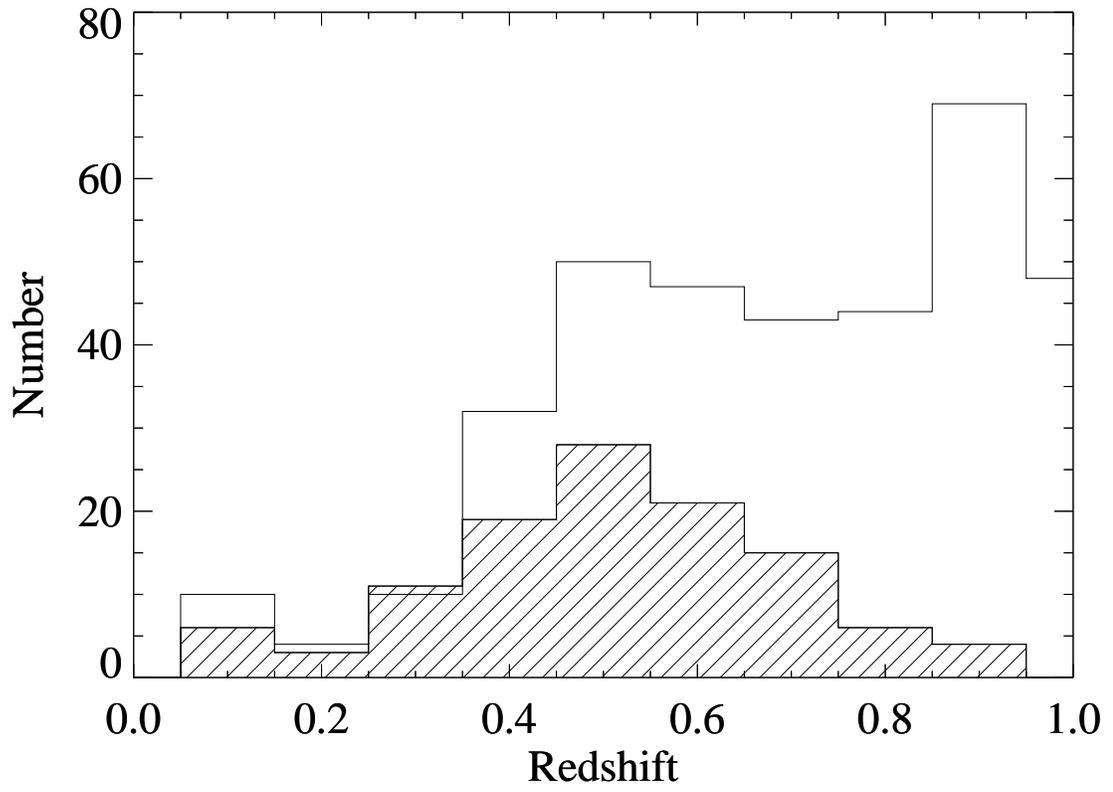}
\caption{ \label{fig: zhist} Distribution of redshifts.  We plot the distribution
  of redshifts for NICMOS-selected sources in the HDF-N (solid line) and
  FUV-detections (filled histogram).}
\end{figure}

Published spectroscopic redshifts are available for 60 of the 135 FUV
detected sources \citep{Cohen 2000, Cohen 2001, Dawson 2001}.  One
spectrum shows the object to be a star.  For the remaining objects,
photometric redshifts have been estimated based on WFPC2
$U_{300}B_{450}V_{606}I_{814}$, NICMOS $J_{110}H_{160}$, and
ground-based $Ks$ \citep{Budavari 2000}.  There were 12 objects that
did not have NICMOS identifications because they were either too faint
or were incorrectly associated with other galaxies/stars.  Two objects
have published photometric redshifts at $z>1$, which would place the
FUV filter blueward of the 912$\AA$ break.  One object has $z_{phot} =
1.09$ so can easily be at $z<1$.  The second source is at $z_{phot} =
2.18$ but appears to be coincident with a background object with
different optical colors.  Figure \ref{fig: zhist} show the
distribution of redshifts.

In the following analyses, we have removed 10 of the 135 sources for
various reasons: the object lies on the edge of the UV image (4 sources), the
object lies on the edge of the NICMOS image (1 source), either the NICMOS or
WFPC2 apertures encompass more than one source and therefore have
compromised photometric redshifts (3 sources), or the object is a star (1).
We also exclude an elliptical at $z=0.089$ from analysis
involving FUV to optical colors, because the FUV light is centered on
a very small region compared to the very large aperture containing the
light in the F300W and other filters; this difference, combined with
the low redshift will result in the F300W being dominated by light
from older stars at the red end of the filter.  These objects and their
fluxes are included in \ref{tbl-1} with footnotes denoting the object
specific problem.

\section{Discussion}
\label{sec: discussion}
\subsection{Number Counts}

\begin{figure}[t*]
\plotone{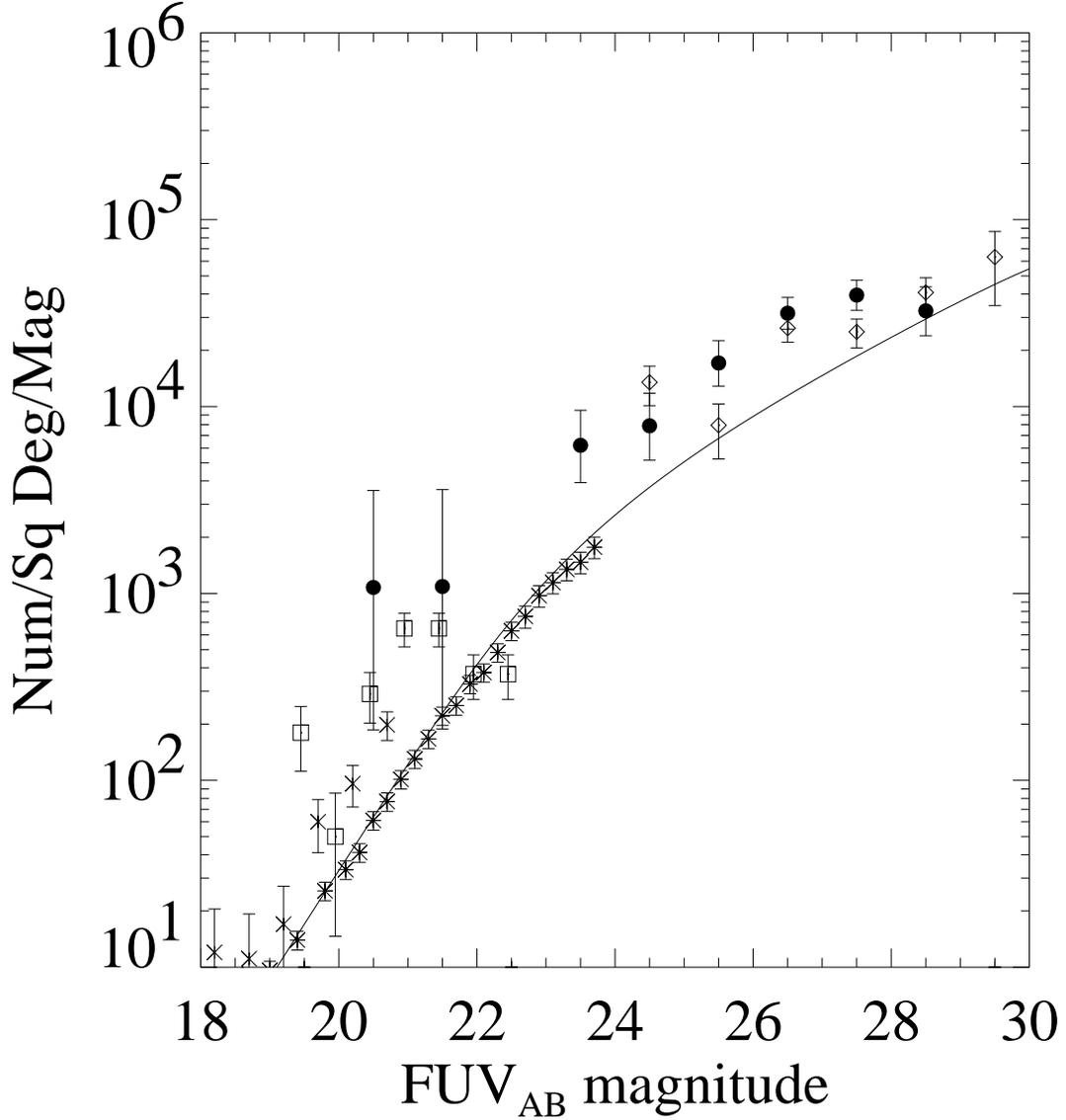}
\caption{ \label{fig: nc} FUV number counts.  We plot the galaxy number-magnitude counts
  for the ACS (solid symbols) and STIS \citep[open symbols; ]{Gardner
    2000 GBF}\ sources.  The STIS counts include seven fields of view
  near the HDF-North and a single pointing in the HDF-South data.  We
  plot for comparison the XMM \citep[squares;][]{Sasseen 2002}\ and
  FOCA \citep[X's;][]{Milliard 1992}\ counts, corrected from 2000 \AA\ 
  to 1500 \AA\ assuming a slope $\beta=-0.8$.  We also plot the GALEX
  number counts (asterisks) without a color correction for filter
  differences, and their model fit which is closest to the HST counts
  \citep{Xu 2005}.}
\end{figure}

We measure galaxy number-magnitudes for the ACS sources and compare
them to the published STIS counts (Figure \ref{fig: nc}). Dark current
variation complicates the measurement of the counts.  We use the
procedure outlined in \cite{Gardner 2000 GBF}.  First, we use the
variance map to determine the area over which each galaxy would have
been detected.  The STIS and ACS counts are generally consistent.
Only one object was measured in common.

\begin{figure}[t*]
\plotone{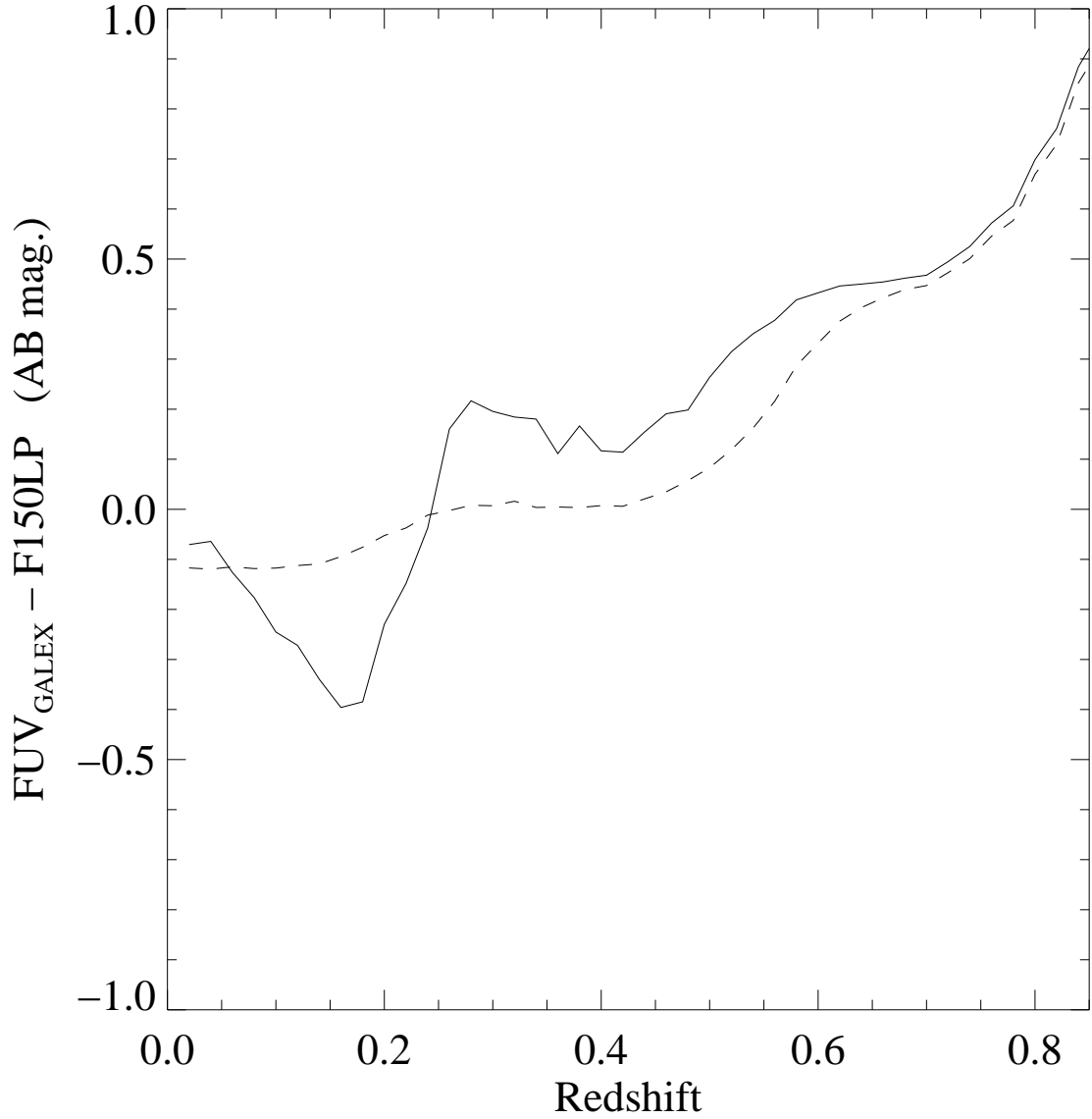}
\caption{ \label{fig: galex_color_corr} The color correction between the GALEX FUV filter
  and the ACS F150LP filter for sources with \lya\ emission (solid
  line) and without it (dashed line).  The color correction was
  calculated based on the template spectra of \cite{Kinney 1993}. }
\end{figure}

In the figure, we also compare the HST number counts to the FUV number
counts measured by GALEX \citep{Xu 2005}, XMM \citep{Sasseen 2002},
and FOCA \citep{Milliard 1992}.  These other counts reach measurement
reached $FUV_{AB}\le 24$.  We correct the XMM and FOCA counts from
their central wavelength of 2000 \AA\ to 1500 \AA\ following Xu et
al.\ by assuming a UV slope, $\beta=-0.8$.  We do not include a color
correction for the difference between the ACS and GALEX filters, but
we estimate that such a correction could be substantial for distant
sources (see Figure \ref{fig: galex_color_corr}).  The difference for
$z>0.5$\ is the result of the bluer wavelength coverage of the GALEX
filter, which is more strongly affected by the 912 \AA\ limit (see
Figure \ref{fig: filter_curve}). The redshifting of the Lyman limit
combined with the redder transmission of the ACS filter causes it to
be sensitive to a larger volume than the GALEX filter, by a factor of
$\sim 30$\%.  For $z\sim 0.15$\ sources, the color correction is
reversed for sources with strong \lya\ emission lines falling in the
GALEX filter but below the blue end of the F150LP filter.  About half
of the ACS sources lie at $z>0.5$.  The color correction would be less
extreme for the STIS filter, which had a red-end cutoff between that
of GALEX and ACS.

The HST counts are higher (by a factor of $\sim 2$) than both the
GALEX counts and the model which fits them.  At $FUV_{AB}\le 24$, the
descrepancy is only marginally significant as the ACS counts lie
within 1-2 $\sigma$\ of the the GALEX counts.  The XMM and FOCA counts
are also higher than GALEX at these magnitudes.  The difference with the
model is more significant, as it is repeated over a larger number of bins.

The GALEX counts are fit by a model which assumes essentially pure
luminosity evolution, $L_*\sim (1+z)^{2.5}$, and a starburst SED that
is flat between 1000 and 1200 \AA\ \citep{Xu 2005}.  The HST counts
are significantly higher than the model.  Some of this difference is
the result of the filter difference discussed above, which causes a
$\sim 30$\% difference in the volume surveyed and potentially a half
magnitude of color-correction.  Thus, there cannot be much more than a
factor of $\sim 2$\ between the counts and the model. The most likely
explanation for this difference is cosmic variance.  The HST counts
are dominated by the very small field of view of the HDF-N, as the
STIS counts include only a single pointing in the HDF-S and seven in
the HDF-N.  The northern sightline is known to have source
overdensities at $z\sim 0.45$\ and $z\sim 0.8$\ \citep{Cohen 2000}.
So, it may not be surprising that the HST counts are higher.
\cite{Somerville 2004}\ estimate that in a typical area the size of
the HDF, the cosmic variance of highly clustered sources is a factor
of $\sim 2$.  We also note that the higher HST counts could indicate
that the pure luminosity evolution model is not sufficient at the
faintest UV fluxes, and perhaps number density evolution is required
as well.  

\cite{Gardner 2000 GBF}\ report that the FUV number counts measured from
the STIS subset of the data are surprisingly flat compared to the
predicted counts \citep{Granato 2000}.  We see no significant change
in the slope with the addition of the ACS data.
  
Finally, we also examined the HST data to determine whether the
difference in spatial resolution between GALEX and HST could result in
source confusion.  There is no evidence of confusion in the HDF at the
depth of the \cite{Xu 2005}\ counts, $FUV_{AB}<24$.  However, at the
depth of the GALEX ultradeep surveys, $FUV_{AB} \sim 26$, some
individual sources would be confused.  The 34 sources brighter than 26
in the ACS area would correspond to $\sim 20$\ beams per source at the
GALEX resolution.


\subsection{Star Formation Rates and Comparison to Other Wavelengths}
\label{sec: SFR}

The detection of HDF-N sources in the FUV provides a sample of
starbursts, and other star-forming galaxies, out to redshifts near unity.  We can
compare their FUV properties to the extensive data available in the
field outside of the HST wavelength range.  

Strong starbursts should also be mid-infrared (MIR) bright galaxies.
UV light absorbed by dust is re-radiated in the far-IR; and heated
dust grains themselves, both small grains and polycyclic aromatic
hydrocarbons (PAHs), emit in the mid-IR.  The ``Great Observatories
Origins Deep Survey'' (GOODS; Dickinson et al.\ 2005, in prep.)
Spitzer Legacy Program has obtained ultra-deep observations of the
field with the Infrared Array Camera \citep[IRAC;][]{Fazio 2004} and
the 24 $\mu$m array of the far-IR photometer \citep[MIPS][]{Rieke
  2004}.  Only 56 of the FUV sources are detected by IRAC and 18 by
MIPS at $\ge 2\sigma$.  For comparison, we examine the IRAC catalog
for the {\it Chandra}\ Deep Field South (Dickinson et al.\ 2006, in
preparation) with the publically available GALEX
catalog\footnote{http://galex.stsci.edu/GR1}.  We find that the
average FUV-IRAC1 color is 2.1 magnitudes.  The HDF FUV image reaches
$AB \ge 29$ while the IRAC channel 1 image reaches $AB \sim 25$\ 
(completeness limit due to confusion), so UV-luminous objects with
typical colors will be more easily detected in the FUV.

\begin{figure}[t*]
\plotone{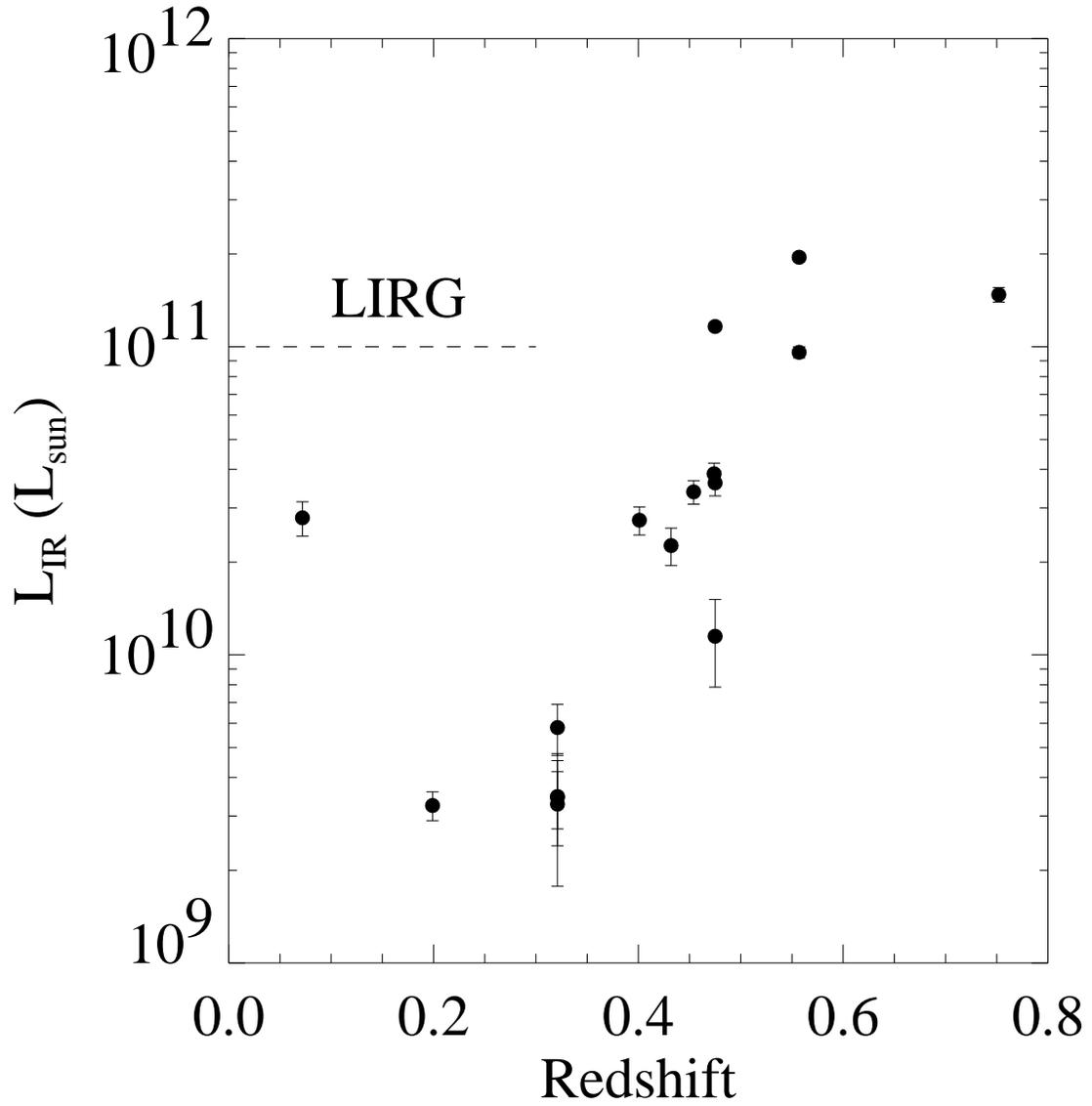}
\caption{ \label{fig: lirz} The inferred \lir\ of MIPS counterparts to FUV sources.
  The errorbars include MIPS photometric uncertainty and a systematic
  term estimated from the difference between \cite{Chary and Elbaz
    2001}\ and \cite{Dale 2002}\ templates.}
\end{figure}

The MIR luminosity of local galaxies in the IRAS bright galaxy sample
\citep{Soifer 1987}\ has been found to correlate strongly with their
far-infrared luminosity which is dominated by large, cool dust grains
\citep{Chary and Elbaz 2001}.  This correlation has been applied to
develop a library of model templates of the mid- and far-infrared SED
of galaxies.  The library consists of template SEDs across a range of
luminosities, which can be redshifted to predict the MIR flux of a source
with given luminosity at a redshift of interest.
For each source in the FUV sample, we select the template for which
the library predicts the closest 24 \mic\ flux density at the
appropriate redshift to apply a bolometric correction; we do not use
the shorter wavelength IRAC measurements.  The corrections based on
the \cite{Chary and Elbaz 2001}\ and \cite{Dale 2002}\ template are
used to derive an infrared luminosity (\lir$=8-1000$\ \mic).  This
technique of deriving the bolometric luminosity from the rest-frame
MIR luminosity is shown to be accurate to 40\%\ in the local Universe
\citep{Chary and Elbaz 2001}.  The difference between the derived
\lir\ from the two templates is assumed to be representative of the
systematic uncertainty in the bolometric correction.  Statistical
uncertainties are assumed to correspond to the signal to noise ratio
of the source at 24 \mic.  The validity of the mid- to far-infrared
correlation and the one-to-one correlation between the bolometric
correction and the MIR luminosity has been tested for field galaxy
samples out to $z\sim 1$\ \citep{Appleton 2004, Marcillac 2005}.  The
MIPS detected sources have a median luminosity of $\sim
10^{10}~L_{\odot}$ with four sources falling in the class of luminous
infrared galaxies (LIRGs; \lir$\simgt 10^{11}~ L_{\odot}$).  Figure
\ref{fig: lirz}\ shows the inferred luminosity of FUV sources detected
by MIPS.

\cite{Meurer 1999}\ find a correlation between the the UV slope
$\beta$ and ratio of infrared luminosity to UV luminosity for
starburst galaxies, where $f_{\lambda} \propto \lambda^{\beta}$. We
measure this slope by combining the FUV data with WFPC2 photometry in
order to measure the slope of the UV continuum from the available
data.  We estimate the value of $\beta$\ following the technique used
by \cite{Meurer 1999}. We begin with 17 spectra from the catalog of
\cite{Kinney 1993}, spanning a range in $\beta$\ as measured by Meurer
et al.  After ``redshifting'' each spectrum from $z=0$\ to $z=0.85$\ 
in steps of $\delta z=0.1$, we integrate under the filter transmission
curves for the ACS F150LP, and the WFPC2 F300W and F450W at each
redshift.  For each redshift, this allows us to define a linear
relation between the color of the object and its intrinsic FUV slope,
yielding a function for $\beta(z, \mbox{color})$.  At low redshift we
use the F150LP-F300W color.  At $z>0.4$, the F150LP filter contains
little information redward of restframe 1000 \AA, so we use the
F300W--F450W color.  We estimate the error in $\beta$\ to be $\sim
30$\%\ from the photometric uncertainty combined with the formal error
in the linear fit to the template values.  In the redshift range
$0.2<z<0.4$\ the F150LP filter includes the \lya\ line, which can
strongly affect the estimate of $\beta$; furthermore, at these redshifts
the FUV filter is stronly affected by the flattening of the
$<1200$\ \AA\ continuum \cite{Buat 2002}.  Thus, we will consider values
for objects at that redshift to be 50\%\ more uncertain; there are
only 6 objects with IRAC counterparts in that range.

We obtain $\beta$\ values with a median of $-1.2\pm 0.6$, with no
clear trend in redshift or FUV magnitude.  \cite{Schiminovich 2005}\ 
measure the UV slope for nearly one thousand galaxies out to $z\sim
1$\ with $FUV_{\mbox{AB}}<24$, using GALEX photometry and VVDS
redshifts.  They report a median slope $\beta_{GLX}=-1.44\pm 1$\ for
sources where confusion is not an issue, in good agreement with other
estimates \citep{Treyer 2005,Adelberger and Steidel 2000}.
 
\begin{figure}[t*]
\plotone{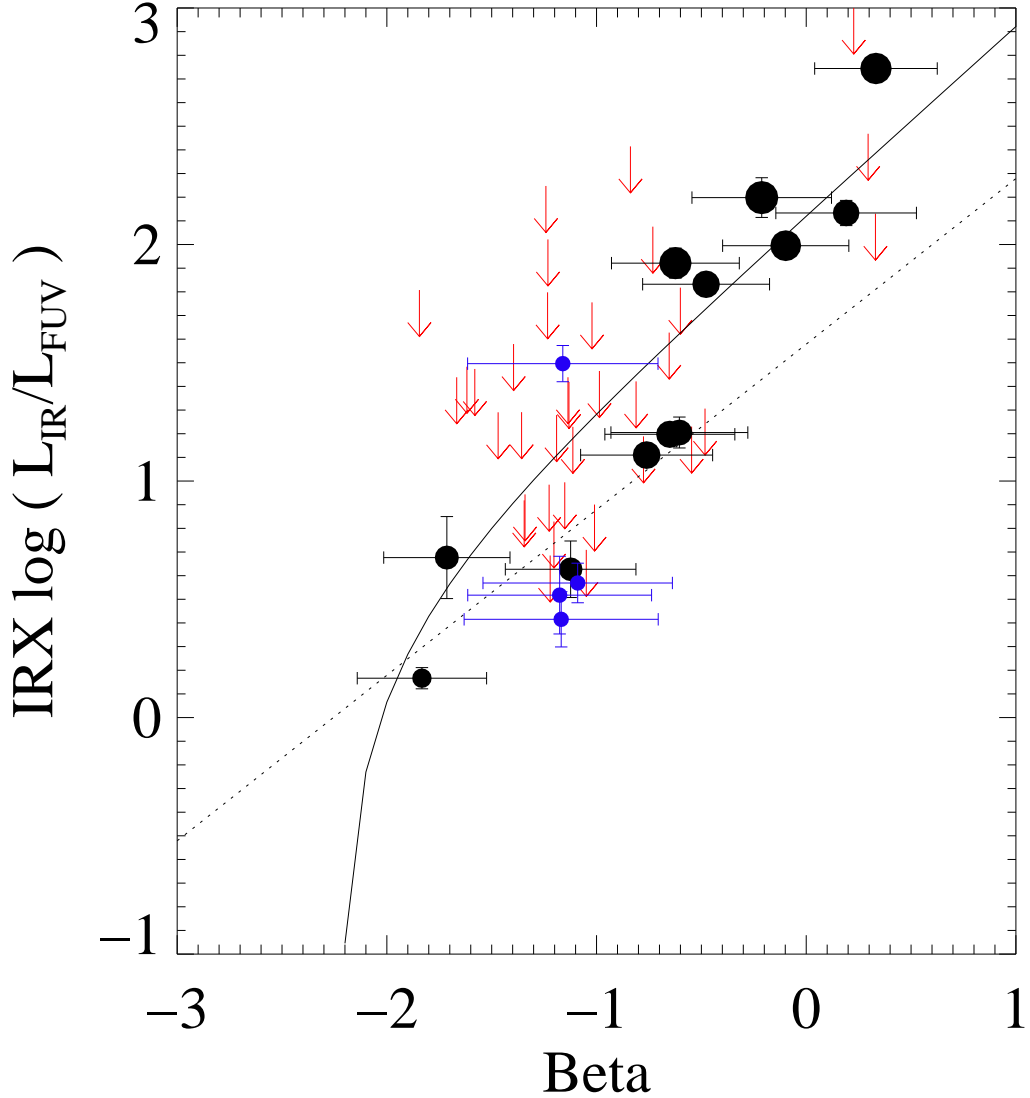}
\caption{ \label{fig: LIR} The ratio of inferred \lir\ vs. UV luminosity ($\nu f_{\nu}$), IRX, is plotted
  against the UV slope, $\beta$.  The symbol size is proportional to
  the inferred \lir\ (larger symbols for more luminous objects).  Two
  $\sigma$\ upper limits are plotted for objects with IRAC
  counterparts but without 2$\sigma$\ MIPS detections.  Error bars
  include statistical and systematic uncertainties as described in the
  text.  Objects at $0.2<z<0.4$\ have higher uncertainty in the
  measurement of $\beta$\ and are plotted in blue.  The solid line
  indicates the the relationship determined by \cite{Meurer 1999}\ and
  the dotted line indicates the relationship measured by \cite{Cortese
    2005} for normal star-forming galaxies.}
\end{figure}

For each source, we also estimate the FUV luminosity ($\nu L_{\nu}$)
at rest-frame 1600 \AA, using the derived value of $\beta$\ and the
flux density in either F150LP or F300W, whichever is closer in the
rest frame.  In Figure \ref{fig: LIR}, we plot the ratio of inferred
\lir\ to $L_{\mbox{FUV}}$ (hereafter $IRX$) against the slope of the
UV continuum and indicate the inferred \lir\ of the sources.  We find
that the relationship of \cite{Meurer 1999}\ is generally reproduced
for the more luminous objects.  Less luminous sources tend to fall
below the line, a trend already noted in GALEX results by
\cite{Seibert 2005}.  In the figure, we also plot the $\beta$\ 
relationship measured by \cite{Cortese 2005}\ for normal star-forming
galaxies which are less luminous than the Kinney et al.\ starbursts.
The less luminous objects in our sample approximately follow the
normal galaxy relation.  Alternately, these sources may have a
contribution from older stars to the filters used in the fit, causing
their $\beta$\ values to appear redder.  This latter explanation may
be less likely, as our sample is FUV-selected, while the Coresse et
al.\ sample is not.  \cite{Burgarella 2005}\ find that the influence of
older stars to the UV colors of galaxies in a UV-selected sample is
small.  At the other extreme, infrared luminous sources have been
observed to generally have higher $IRX$\ for a given $beta$\ than the
starburst galaxies in the Meurer et al.\ sample \citep{Goldader 2002}.
The brightest LIRG in our sample does lie slightly above the line.

\subsubsection{Star Formation Rates}

\begin{figure}[t*]
\plotone{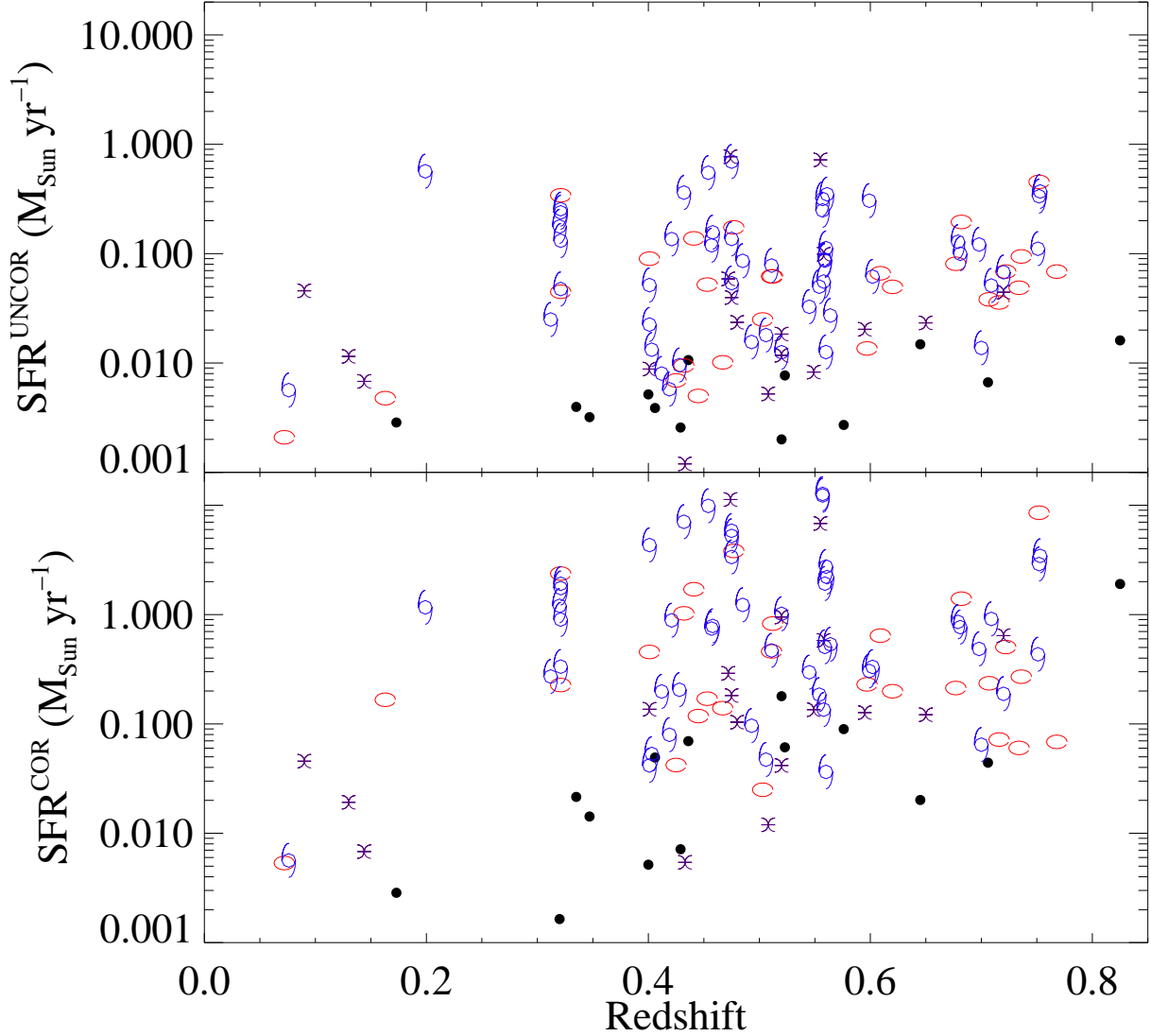}
\caption{ \label{fig: SFR} The inferred star formation rates for FUV detected sources,
  with (bottom) and without (top) extinction correction, $A_{FUV}$.
  Morphological type are indicated by symbols: red ellipses for E/0,
  purple stylized asterisk for peculiar/irregular, blue stylized spiral for later
  than S0, and solid circles for objects too faint to classify.
  Classifications were performed by eye in the rest frame B-band \citep{Conselice 2005} 
  as described in Section \ref{sec: morph}.  }
\end{figure}

FUV imaging provides a powerful tool for measuring the star formation
in normal galaxies, but is strongly affected by extinction.  
We can derive the star formation rate (SFR) for each
galaxy from the detected FUV flux.  \cite{Kennicutt 1998}\  gives the
calibration from the 1500 \AA\ continuum to the SFR:

\begin{displaymath}
\rm{SFR} (M_{\odot}~\rm{yr}^{-1}) = 1.4\times 10^{-28}L_{\rm{FUV}}~(\rm{ergs}~\rm{s}^{-1}\rm{Hz}^{-1}),
\end{displaymath}

\noindent assuming a Salpeter IMF \citep{Salpeter 1955} with mass limits of
0.1-100 \Msun and continuous star formation.  The \cite{Meurer 1999}\ 
IRX relation suggests a calibration for the dereddening factor of
$A_{\rm{FUV}} = 4.43 + 1.99(\beta)$, expressed in magnitudes.  For the
FUV-detected sources, we obtain a median extinction factor, as a
multiple rather than in magnitudes, of $\sim 6$.  The relationship
assumes, however, that the UV flux is entirely the product of young
stars.  Our sample likely includes sources with only moderate star
formation, ordinary spirals and even elliptical galaxies, so the flux
within the filter wavelength range (particularly in the WFPC2 filters)
may include some contribution from an older (or aging) stellar
population.  The star formation rates (SFRs) that we infer will thus
be upper limits.  In Figure \ref{fig: SFR}, we show the inferred star
formation rates for FUV sources as a function of redshift and
morphology.

\subsubsection{X-ray Properties}

The Chandra 2 Ms catalog \citep{Alexander 2003}\ contains 13 sources
within the FUV survey area with spectroscopic redshifts of $z<0.85$\ 
from the catalog of \cite{Barger 2003}.  Six optical counterparts
to these X-ray sources are detected in the FUV.  These objects
lie at redshifts 0.089, 0.139, 0.475, 0.556, and 0.752.  The object
at 0.321 is described by \cite{Barger 2003} as a possible multiple structure
contaminated by a foreground object; the optical/FUV counterpart is more than
an arcsecond from the X-ray position.
Three of the FUV counterparts are spatially extended, and the other
three are extremely faint in the FUV but extended in the F450W filter.
None of them are detected in the hard band, none have broad optical
emission lines \citep{Barger 2003}\ and all are near the detection
limit of the softband (0.5-2 keV), with fluxes $SB \le 0.08$\ ergs
cm$^{-2}$ s$^{-1}$.  These properties are consistent with the
interpretation that the source of the X-rays is star formation rather
than active galactic nuclei.  Similarly, the \cite{Ranalli 2003}\ 
calibration of X-ray luminosity as a star formation indicator yields
rates generally in agreement (within a factor of a few) with the
UV-inferred SFR.

\subsection{Morphological Distributions}
\label{sec: morph}

FUV imaging picks out the location of the most recent star formation.
Photometry tells us the total SFR; morphology tells us where it occurs
within the galaxies.  As a result, the appearance of a galaxy can vary
dramatically in different passbands even in the absence of dust,
\citep[a ``morphological k-correction'', see ][and the references
therein]{Papovich 2003}.  In Figure \ref{fig: morph kcorr}, we compare
the morphologies of selected UV-bright galaxies in the FUV.  As
expected, some galaxies appear similar across wavelengths while others
show substantial differences.  Truly irregular or morphologically
disturbed galaxies tend to appear similar across wavelength, as do
some elliptical galaxies \citep[see counter examples in][]{Windhorst
  2002}.  The morphological K-correction is most pronounced for early
to mid-type spirals, in which a substantial population of old stars
defines the optical shape, but regions of recent star formation are
``lit up'' across the galaxy \citep{Windhorst 2002}.

We avoid the K-correction by matching the FUV detections to a
morphological catalog of galaxy types \citep{Conselice 2005}\ 
and CAS parameters \citep[see ][]{Conselice 2003 CAS}.
The catalog includes galaxy morphologies in the rest-frame B-band
for the 200 HDF galaxies out to $z=0.85$\ that are bright enough for
visual classification (out of 240 possible).  We restrict our analysis
to $z < 0.85$ because the redshifting Lyman break leaves little flux
within the F150LP filter at higher redshift.  We exclude 10 sources
that are not in the \cite{Conselice 2005} catalog, mostly due to the lack of
NICMOS counterparts.

\begin{figure}[h*]
\plotone{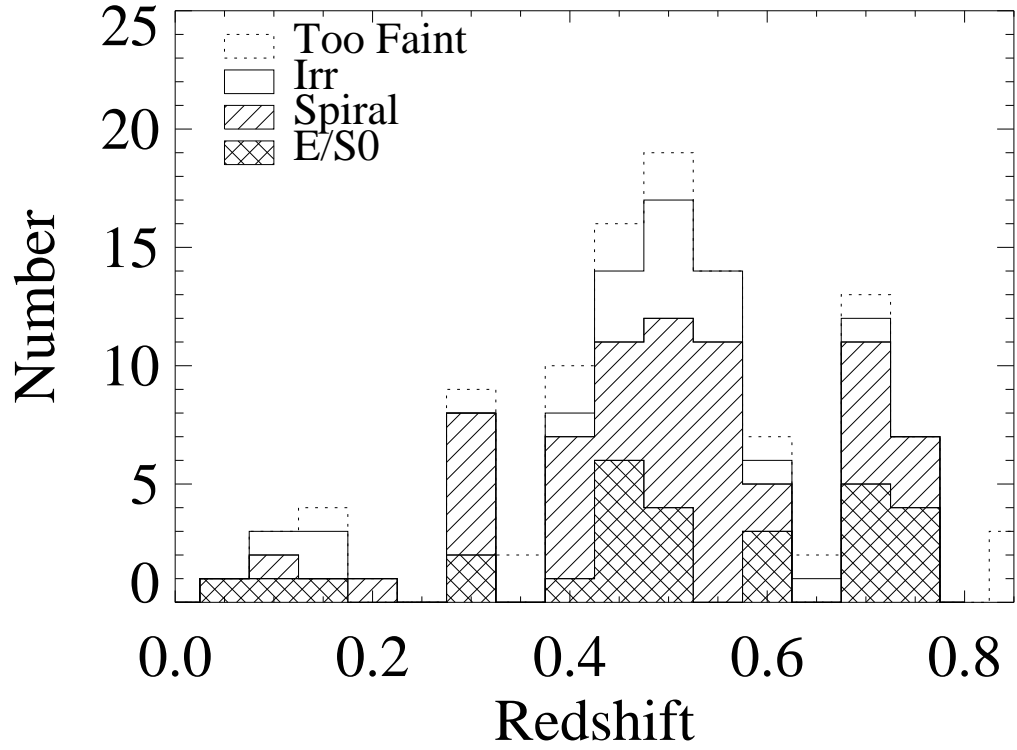}
\caption{ \label{fig: zhist morph} The distribution of redshifts for each 
morphological type.  Classifications were performed by eye in the rest frame
B-band (see text).}
\end{figure}

This gives us some indication of the types of galaxies that are
emitting in the FUV at redshifts $z < 1$. We find that the galaxies
detected in the FUV span all the major morphological types, as also
seen by \cite{de Mello 2004}.  Figure \ref{fig: zhist morph}\ shows
the morphological break down for systems based on their apparent
morphological types as classified by \cite{Conselice 2005}.  As can be
seen the spiral galaxies dominate the number counts for the FUV
sources, although spheroids make up a significant fraction of the
detections at $z > 0.6$ and irregulars are also represented. We find,
in fact, that a significant fraction of all spheroids (30/56) and a
similar fraction of spirals (50/87) at $z < 0.85$\ are detected in the
FUV in the HDF-N.

\begin{figure}[h*]
\plotone{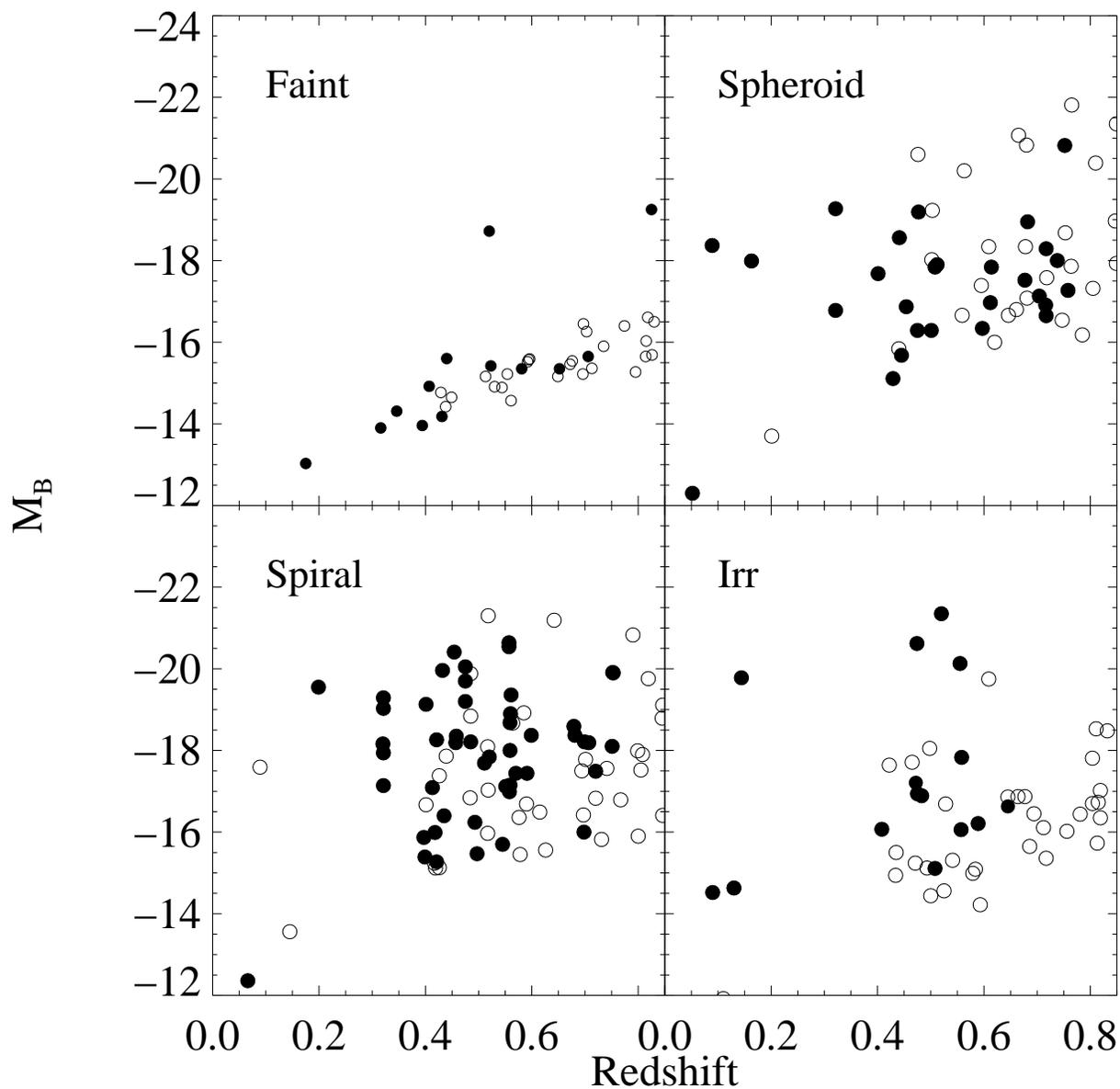}
\caption{ \label{fig: fuv mb} Absolute magnitude as a function of redshift.  We plot
  $z$\ vs. $M_{\rm B}$\ for each morphological type: spheroids (upper
  right), peculiar/irregular (lower right), later than S0 (lower
  left), and objects too faint to classify (upper left).
  Classifications were performed by eye in the rest frame B-band 
  \citep[][see text]{Conselice 2005}.  Objects detected in the FUV are
  plotted with filled circles.  }
\end{figure}

The relative distributions of FUV emitting types with redshifts can be
seen in Figure \ref{fig: fuv mb}, which plots the absolute M$_{\rm
  B}$ magnitude as a function of redshift.  From this diagram, there
is a broad range of absolute magnitudes for the FUV sources at all
redshifts.  The median luminosity for UV-detected spheroids, spirals,
and irregulars is $-17.3$, $-18.2$, and $-16.2$, respectively, compared to
median values of $-17.7$, $-17.8$, and $-16.4$\ for all galaxies of the three
types at $z<0.85$\ in the HDF.  Interestingly, while the median values
agree, the most luminous ellipticals in the HDF (M$_{\rm B}<-19$) are
not generally detected in the FUV (3 out of 11).

The figure shows that less than half of peculiar/irregular at $z<0.85$\ 
are detected in the FUV; only 38\% (18/47) have FUV detections.
This is due in large part to their intrinsic faintness, rather than
unusually red color.  Half of the irregulars have V-band magnitudes
fainter than 27, which makes them undetectable in some or all areas of the
FUV image. The FUV-detected irregulars are somewhat bluer
than the rest of the UV sample, with a median value of
$FUV_{AB}-V_{AB}\sim 1.3$\ compared to a median color of 1.5 for the
entire FUV catalog.  These same objects are optically blue, with a
median $V-I\sim 0.5$\ compared to 0.6 for the all $z<0.85$\ sources in
the HDF.

\subsection{Star formation in Spheroids}

\begin{figure}[t*]
\plotone{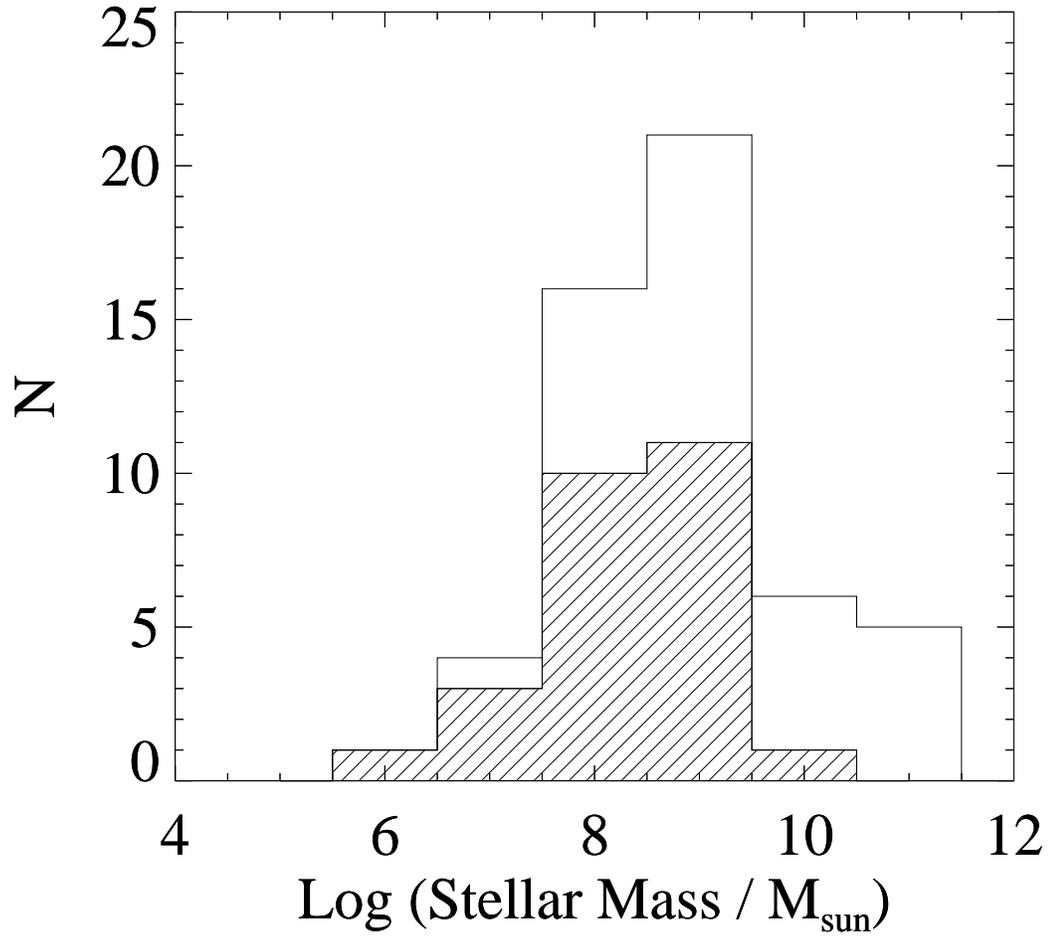}
\caption{ \label{fig: massdisp} Histogram of distribution of inferred mass for spheroids in the HDF.  Sources
detected in the FUV are shown by the hatched histogram.  }
\end{figure}

We find evidence for star formation in $\sim 50$\%\ of spheroids at
$z<0.85$.  These objects are typically less massive than $10^{10}$\ 
\Msun\ (see Figure \ref{fig: massdisp}) and less luminous than
$M_{B}=-19$.  Their sizes (half light radii) are similar to other spheroids
in the HDF.  We find a median SFR of 0.25 $M_{\odot}$\ yr$^{-1}$,
after extinction correction (see Section \ref{sec: SFR}).

So far, we have only considered morphologically-selected spheroids.
However, few of these objects have the SED of purely old stellar
populations, even without including the FUV.  \cite{Stanford 2004}\ 
show that morphologically- and spectroscopically-identified spheroids in
the HDF are not necessarily the same population.  A morphological
selection identifies sources which have SEDs similar to local
spheroids and identifies additional sources that are bluer, less
massive, and less luminous than those.  Only one of the objects in the
\cite{Stanford 2004}\ sample of spectroscopic ellipticals is detected
in the FUV.  Their sample of morphologically selected ellipticals is
not identical to that of \cite{Conselice 2005}, but the fraction of
FUV detections is similar.  The FUV detection, then, supports the
\cite{Stanford 2004}\ conclusion that some morphological spheroids have
recent or ongoing star formation.

\begin{figure}[t*]
\plotone{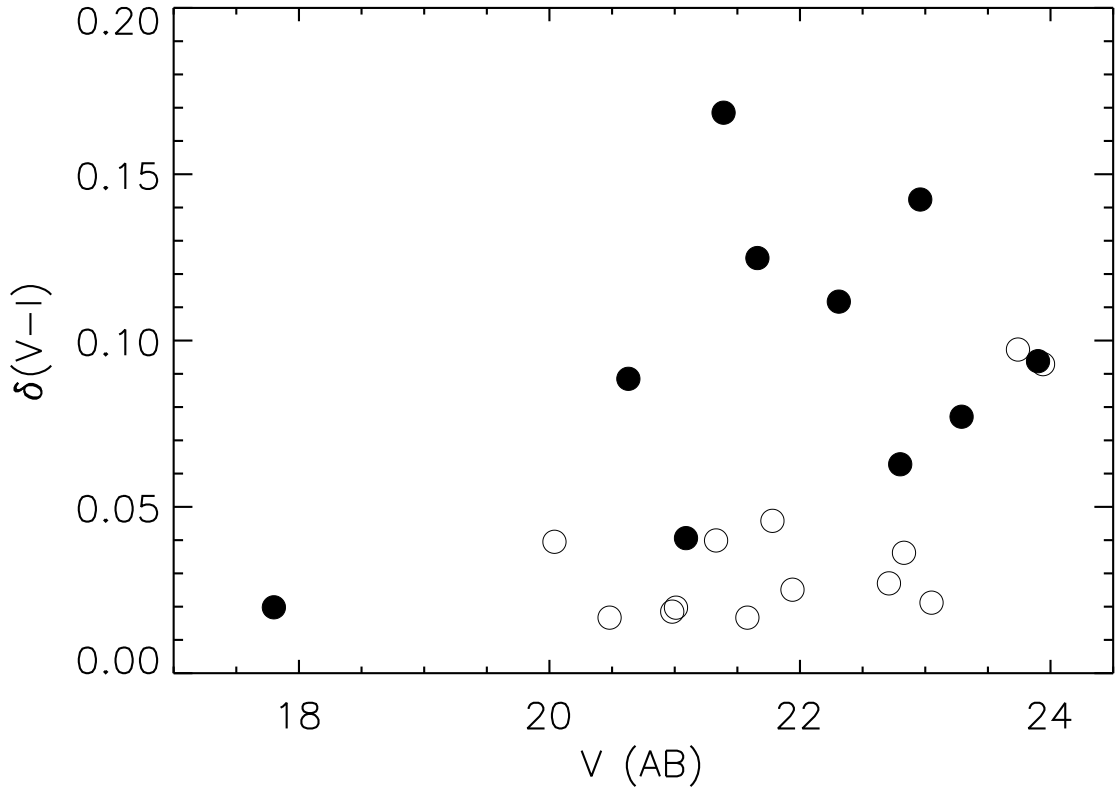}
\caption{ \label{fig: blue-core} FUV detection of blue core ellipticals.  We plot the color 
  gradient for ellipticals identified in \cite{Menanteau 2001}\ vs.
  their $F450W$\ magnitidue.  A larger color gradient indicates a
  bluer object.  Filled symbols indicate FUV detection.  }
\end{figure}

This effect has also been seen in the ``blue-core'' ellipticals
\citep{Menanteau 2001}.  These objects were initially identified by
strong color gradients in the WFPC2 images, which show significant
bluing towards the center.  The presence of the blue cores suggested a
population of young ($<1$Gyr) ellipticals which may have undergone
recent merger activity or some type of residual star formation.
Ten of the 21 sources at $z<0.85$\ in the Menanteau et al.\ sample
are detected in the FUV.  In Figure \ref{fig: blue-core}, we show that
the sources with the bluest cores are the ones most likely to be
detected in the FUV.  The detection of FUV flux near the core of the
sources confirms that these objects have small amounts of ongoing star
formation.

It is highly unlikely that the FUV flux detected in spheroids is
the result of the ``UV upturn'' that arises from a minority population
of hot horizontal branch stars \cite[e.g.][]{Brown 1997}.  For
example, cluster ellipticals at $z= 0.33$\ and $z=0.55$\ have been
observed to show small amounts of UV emission \citep{Brown 2000b,
  Brown 2003}.  Unlike the HDF spheroids, these objects have optical
SEDs broadly consistent with old stellar populations.  The UV emission
in UV-upturn galaxies is a small fraction of the total luminosity.
\cite{Brown 2003}\ find $m_{1500}-V\sim 4$\ for UV-upturn galaxies,
while the median color of the HDF spheroids is $\sim 2$\ mag.\ and
only one of them has a color greater than 3.  The differences in the
STIS and ACS filters could account for ACS objects being $\sim 0.4$\ 
magnitudes brighter than their STIS counterparts at $z>0.2$, but the
HDF spheroids are still significantly brighter in the UV.  Similarly,
\cite{Brown 2003}\ estimate that the flux associated with the UV
upturn would correspond to $SFR \sim 0.005-.02$\ \Myr\ if it arose
instead from star formation.  The inferred SFR in HDF spheroids is a
factor of several higher even without extinction correction, and
significantly higher with the correction.  However, we note that an
old stellar contamination of the redder filters could result in an
overestimate of the SFR.  Thus it is likely, though not certain, that
most of the UV flux in HDF spheroids is the result of star formation
and not the UV upturn.

The more massive and luminous ellipticals in the HDF appear not to be
forming stars at rates similar to the smaller and fainter ones that we
detect in the FUV.  This could be a direct indication that lower
luminosity ellipticals in the field form later than the giant
ellipticals.  This is consistent with the widely varying ages measured
for local ellipticals \citep{Trager 2000}.  Such a distinction may be
evidence of downsizing in the galaxy formation process, which may be
directly related to the rate of merging which is seen to be high for
lower luminosity and lower mass galaxies at $z < 1$ \citep{Conselice
  2003}.  

\begin{figure}[t*]
\plotone{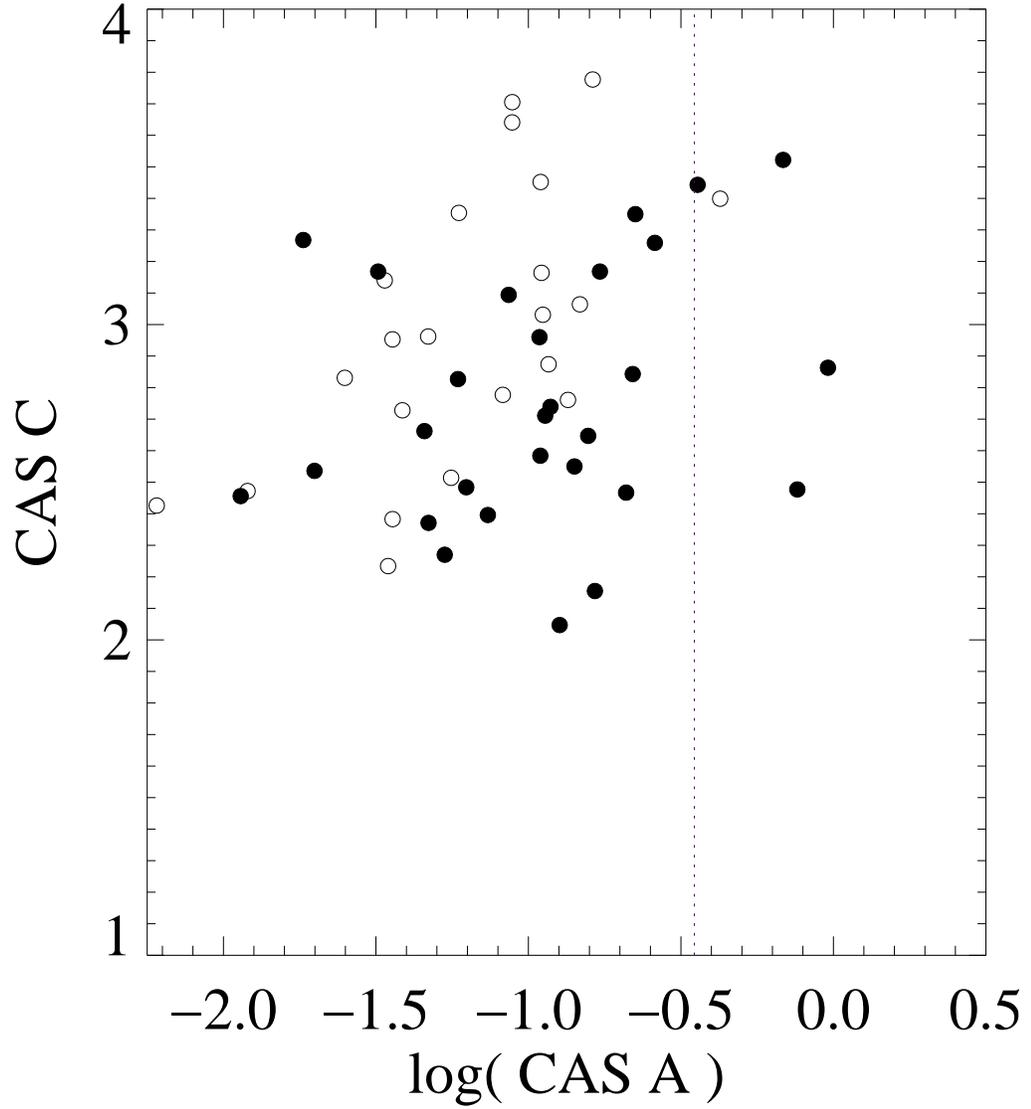}
\caption{ \label{fig: CAS} The concentration and asymmetry plane for 
  spheroids at $z<0.85$\ in the HDF \citep{Conselice 2003}.  The morphology has
  been measured in the filter corresponding to rest-frame $B$-band. Sources
  detected in the FUV are plotted as filled circles.  The dotted line
  indicates indicates the minimum asymmetry for typical mergers.  }
\end{figure}

Another way to investigate the star-formation nature of early type
galaxies is by examining their location in the concentration-asymmetry
diagram for galaxies at $z < 1$ (Figure \ref{fig: CAS}).  The most
evolved spheroids which have had no star formation in the recent past,
should contain a high concentration and a low asymmetry.  These
objects typically do not have FUV emission. On the other hand,
morphologically-identifiable spheroids with high asymmetries, that
indicate a recent evolution, are more likely than not to have FUV
emission.  This result supports the conclusion that the FUV emission
is originating from star formation, which produces the structural
asymmetries. Furthermore, many of the FUV-detected spheroids have
relatively low concentrations, consistent with their morphology
tracing the regions of young stars as well as the underlying older
population \citep[e.g.][]{Windhorst 2002}.

The inferred SFR in HDF spheroids is not high enough for them to be
the progenitors of local giant ellipticals, but it may suggest that a
significant fraction of the stars in lower luminosity and lower mass
ellipticals form at $z < 1$.  There is no evidence that the FUV
detected spheroids are undergoing the last years of a final episode of
star formation.  Instead, we might estimate a duty cycle of star
formation episodes.  The detection fraction ($\sim 50$\%) suggests
that these objects could spend as much as half the time producing
small amounts of stars.  With a median SFR of 0.3 \Myr, this duty
cycle would allow as much as $\sim 10^9$\ \Msun\ to form between
$z=0.85$\ and $z=0$.  The median stellar mass estimated by
\cite{Conselice 2005}\ for the spheroids is $\sim 10^{8.5}$\ \Msun, so
they could double or triple in size by present day.  They would still
remain much less massive than the spectroscopically-identified
ellipticals, which typically have stellar masses greater than
$10^{10}$\ \Msun\ \citep{Stanford 2004}.

\section{Summary}

We have obtained FUV imaging of the Hubble Deep Field North using the
Solar Blind Channel (SBC) of the ACS and FUV MAMA of the STIS.  We achieve
$3.5\sigma$\ sensitivities fainter than $FUV_{AB}\sim 29$.  We detect
134 galaxies and one star.  We have compared our results to the
multiwavelength data available for the field.  We find the following:

1. The enhanced dark current ``glow'' in the center of the SBC chip
  is a strong function of detector temperature, which rises sharply
  during observation visits longer than two orbits.  Future large SBC
  programs would benefit from breaking observations into multiple
  short visits.

2. Galaxy number-magnitude counts for the full survey generally
  agree with those previously published for a subset of the data, but
  are a factor of $\sim 2$\ higher than a model fit to counts measured
  by GALEX at the brighter magnitudes.  We attribute the difference to
  a combination of: a) differences in the FUV filter transmission
  between the two observatories, and b) cosmic variation resulting
  from the small field of view of the HDF-N.  We see no evidence for
  source confusion at the level of the current GALEX counts,
  $FUV_{AB}\sim 24$, but find that confusion may be an issue in the
  ultradeep GALEX survey at $FUV_{AB}\sim 26$.
  
3. We detect the optical counterparts \citep[as identified by
  ][]{Barger 2003}\ to six of 13 {\it Chandra}\ sources in the field.
  The FUV and X-ray properties of these sources are consistent with
  star formation rather than active galactic nuclei.
  
4. Eighteen FUV-detected galaxies are also detected in the GOODS
  MIPS 24 \mic\ image of the field.  The inferred ratio of infrared to
  ultraviolet luminosities, $IRX$, generally follows the relationship
  with UV-slope, $\beta$, measured for either starbursts \cite{Meurer
    1999}\ or normal galaxies \cite{Cortese 2005}.  Using the
  $IRX-\beta$\ relation to correct for extinction, we infer star
  formation rates of a few tenths of a solar mass per year up to
  almost 10, for the entire sample of FUV detected sources.  The
  median SFR is 0.3 \Myr\ and 75\%\ of sources have SFR$<1$\ \Myr.
  
  5. Rest-frame $B$-band morphologies are available in the literature
  for most of the FUV-detected sources.  Half of the FUV-detected
  sources have spiral morphologies.  We detect only $\sim 40$\%\ of
  galaxies with irregular morphologies, which we attribute to their
  intrinsic faintness rather than unusually red color.  We find
  evidence for star formation in $\sim 50$\%\ of the morphologically
  identified, moderate-mass spheroids at $z<0.85$. These sources
  include the ``blue-core'' ellipticals with the strongest
  color gradients.  As noted by \cite{Stanford 2004}, the
  morphologically-identified spheroids include sources with the SED of
  local ellipticals and other, bluer galaxies.  The former group are
  generally not detected by our survey.  Thus the SED of the spheroids
  supports our identification of the FUV flux as arising from ongoing
  star formation.  The large fraction of FUV-detected spheroids
  suggests they continue to build stellar mass after $z\sim 1$, which
  is supported by their morphological asymmetries.

The small area of the HDF limits the results that can be drawn from
the present survey.  We have undertaken a complementary survey to
obtain FUV imaging of the Hubble Ultra Deep Field \citep{Beckwith
  2003}, and results will be presented in a future paper.  The
combination of ultradeep HST and GALEX imaging of the same fields will
augment the interpretation of both.

\acknowledgements

The research described in this paper was carried out, in part, by the
Jet Propulsion Laboratory, California Institute of Technology, and was
sponsored by the National Aeronautics and Space Administration.
Support for proposal 9478 was provided by NASA through a grant from
STScI, which is operated by AURA, Inc., under NASA contract NAS
5-26555.

\clearpage

\begin{deluxetable}{l r r r r r l}
  \tabletypesize{\scriptsize} \tablecaption{Photometry\label{tbl-1}}
  \tablehead{ \colhead{Object} & \colhead{R.A.\tablenotemark{a}} &
    \colhead{Dec\tablenotemark{a}} & \colhead{Inst \tablenotemark{b}}
    & \colhead{FUV$_{AB}$} & \colhead{$\sigma_{FUV}$} &
    \colhead{HDF ID \tablenotemark{c}} \\
    \colhead{} & \colhead{(J2000)} & \colhead{(J2000)} & \colhead{} &
    \colhead{(mag.)} & \colhead{(mag.)} & \colhead{} } \startdata
  1 &    12:36:39.77 &    62:12:28.75 &    S &    25.98 &     0.16 & 4-852.0         \\
  2 &    12:36:39.87 &    62:12:31.61 &    S &    27.93 &     0.36 & 4-823.0         \\
  3 &    12:36:40.05 &    62:12:21.43 &    S &    27.00 &     0.22 & 4-860.1         \\
  4 &    12:36:40.09 &    62:12:22.24 &    S &    26.57 &     0.14 & 4-860.0         \\
  5 &    12:36:41.15 &    62:12:10.59 &    S &    27.97 &     0.34 & 4-822.0         \\
  6 &    12:36:41.95 &    62:12: 5.40 &    S &    24.62 &     0.09 & 4-795.0         \\
  7 &    12:36:42.92 &    62:12:16.37 &    S &    24.19 &     0.04 & 4-656.0         \\
  8 &    12:36:43.40 &    62:13: 4.76 &    A &    27.66 &     0.24 & 1-43.0          \\
  9 &    12:36:43.41 &    62:11:49.27 &    S &    28.26 &     0.53 & 4-728.0         \\
  10 &    12:36:43.63 &    62:12:18.24 &    S &    27.70 &     0.45 & 4-565.0         \\
  11 &    12:36:43.82 &    62:12:22.41 &    S &    28.26 &     0.31 & 4-525.0         \\
  12 &    12:36:43.98 &    62:12:49.92 &    A &    26.45 &     0.26 & 4-402.31        \\
  13 &    12:36:44.18 &    62:12:47.78 &    A &    24.15 &     0.03 & 4-402.0         \\
  14 &    12:36:44.47 &    62:13: 7.63 &    A &    27.98 &     0.16 & 1-41.0          \\
  15 &    12:36:44.48 &    62:11:53.26 &    S &    27.86 &     0.30 & 4-627.0         \\
  16 &    12:36:44.62 &    62:13:18.94 &    A &    27.66 &     0.16 & 1-76.0          \\
  17 &    12:36:44.70 &    62:13: 6.74 &    A &    27.52 &     0.16 & 1-37.2          \\
  18 &    12:36:44.73 &    62:11:43.81 &    S &    26.81 &     0.16 & 4-658.0         \\
  19 &    12:36:44.74 &    62:11:57.06 &    S &    26.65 &     0.12 & 4-579.0         \\
  20 &    12:36:44.82 &    62:13:17.57 &    A &    27.80 &     0.16 & 1-68.0          \\
  21 &    12:36:44.83 &    62:12: 0.25 &    S &    25.32 &     0.08 & 4-558.0         \\
  22 &    12:36:45.31 &    62:11:42.91 &    S &    26.17 &     0.10 & 4-618.0         \\
  23\tablenotemark{h} &    12:36:45.42 &    62:12:13.55 &    S &    23.60 &     0.03 & 4-454.0         \\
  24 &    12:36:45.43 &    62:13:26.01 &    A &    24.76 &     0.04 & 1-86.0          \\
  25\tablenotemark{d} &    12:36:45.47 &    62:13:56.99 &    A &    27.54 &     0.15 & 2-126.0         \\
  26 &    12:36:45.51 &    62:13:44.14 &    A &    26.41 &     0.10 & 2-62.0          \\
  27 &    12:36:45.54 &    62:13:29.95 &    A &    27.06 &     0.12 & 1-100.0         \\
  28 &    12:36:45.63 &    62:13: 8.89 &    A &    26.26 &     0.08 & 1-35.0          \\
  29 &    12:36:45.86 &    62:13:25.82 &    A &    23.80 &     0.03 & 1-87.0          \\
  30 &    12:36:45.91 &    62:13:44.82 &    A &    26.69 &     0.10 & 2-100.0         \\
  31 &    12:36:45.96 &    62:12: 1.41 &    S &    27.42 &     0.23 & 4-460.0         \\
  32 &    12:36:46.12 &    62:13:34.71 &    A &    27.43 &     0.17 & 2-61.0          \\
  33 &    12:36:46.16 &    62:13:13.93 &    A &    26.69 &     0.09 & 1-47.0          \\
  34 &    12:36:46.36 &    62:14: 4.99 &    A &    28.25 &     1.19 & 2-251.0         \\
  35 &    12:36:46.55 &    62:14: 7.60 &    A &    25.40 &     0.05 & 2-270.0         \\
  36 &    12:36:46.55 &    62:12: 3.10 &    S &    26.02 &     0.09 & 4-416.0         \\
  37 &    12:36:46.58 &    62:11:57.16 &    S &    26.87 &     0.15 & 4-434.0         \\
  38 &    12:36:46.75 &    62:13:12.31 &    A &    26.08 &     0.09 & 2-7.0           \\
  39 &    12:36:46.95 &    62:12: 5.37 &    S &    27.46 &     0.24 & 4-382.0         \\
  40 &    12:36:46.96 &    62:13:27.84 &    A &    26.31 &     0.15 & 2-108.0         \\
  41 &    12:36:47.02 &    62:13:52.99 &    A &    26.55 &     0.09 & 2-231.0         \\
  42 &    12:36:47.05 &    62:12:36.87 &    A &    23.07 &     0.02 & 4-241.1         \\
  43 &    12:36:47.08 &    62:12:12.54 &    S &    27.17 &     0.18 & 4-332.0         \\
  44 &    12:36:47.15 &    62:14:15.96 &    A &    27.28 &     0.12 & 2-354.0         \\
  45 &    12:36:47.23 &    62:11:58.96 &    S &    27.55 &     0.21 & 4-385.0         \\
  46 &    12:36:47.25 &    62:12:12.66 &    A &    27.34 &     0.11 & 4-332.2         \\
  47 &    12:36:47.28 &    62:12:30.81 &    A &    24.69 &     0.04 & 4-232.0         \\
  48 &    12:36:47.41 &    62:14: 3.05 &    A &    25.69 &     0.06 & 2-321.1         \\
  49 &    12:36:47.54 &    62:12:52.68 &    A &    27.13 &     0.13 & 4-89.0          \\
  50 &    12:36:47.73 &    62:13:14.39 &    A &    28.77 &     0.22 & 2-88.0          \\
  51 &    12:36:47.84 &    62:13: 6.48 &    A &    28.25 &     0.23 & 2-121.2         \\
  52 &    12:36:47.89 &    62:12:29.49 &    A &    28.23 &     0.27 & 4-174.0         \\
  53 &    12:36:47.94 &    62:13:11.08 &    A &    27.89 &     0.16 & 2-121.12        \\
  54 &    12:36:47.98 &    62:13:31.93 &    A &    28.18 &     0.18 & 2-197.0         \\
  55 &    12:36:48.13 &    62:12:14.88 &    A &    26.07 &     0.06 & 4-260.0         \\
  56\tablenotemark{e} &    12:36:48.31 &    62:14:26.45 &    A &    20.95 &     0.01 & 2-537.0         \\
  57 &    12:36:48.63 &    62:12:14.13 &    A &    26.91 &     0.13 & 4-260.2         \\
  58 &    12:36:48.73 &    62:13: 2.48 &    A &    28.73 &     0.34 & 3-51.0          \\
  59 &    12:36:48.78 &    62:13:18.60 &    A &    26.44 &     0.13 & 2-210.0         \\
  60 &    12:36:48.92 &    62:12: 8.02 &    A &    27.80 &     0.21 & 4-203.0         \\
  61 &    12:36:49.00 &    62:12:45.84 &    A &    25.86 &     0.13 & 3-258.0         \\
  62 &    12:36:49.35 &    62:11:54.97 &    A &    28.26 &     0.22 & 4-235.0         \\
  63 &    12:36:49.39 &    62:13:11.27 &    A &    25.22 &     0.05 & 2-264.0         \\
  64\tablenotemark{i} &    12:36:49.45 &    62:13:46.85 &    A &    25.82 &     0.23 & 2-456.0         \\
  65 &    12:36:49.50 &    62:14: 6.69 &    A &    27.12 &     0.17 & 2-514.0         \\
  66 &    12:36:49.59 &    62:14:14.99 &    A &    27.99 &     0.27 & 2-585.2         \\
  67 &    12:36:49.63 &    62:12:57.79 &    A &    25.66 &     0.06 & 3-143.0         \\
  68 &    12:36:49.77 &    62:13:13.03 &    A &    26.39 &     0.12 & 2-264.1         \\
  69 &    12:36:49.89 &    62:12:42.17 &    A &    27.41 &     0.24 & 3-331.0         \\
  70 &    12:36:50.11 &    62:14:28.68 &    A &    28.74 &     0.19 & 2-681.0         \\
  71 &    12:36:50.17 &    62:14:22.16 &    A &    26.81 &     0.09 & 2-645.0         \\
  72 &    12:36:50.23 &    62:14: 7.62 &    A &    26.81 &     0.10 & 2-575.0         \\
  73 &    12:36:50.24 &    62:12:39.55 &    A &    23.41 &     0.02 & 3-386.0         \\
  74 &    12:36:50.29 &    62:12:53.45 &    A &    28.98 &     0.22 & 3-201.0         \\
  75 &    12:36:50.80 &    62:12:21.36 &    A &    26.46 &     0.08 & 3-599.0         \\
  76 &    12:36:50.82 &    62:12: 0.81 &    A &    27.19 &     0.12 & 4-71.0          \\
  77 &    12:36:50.83 &    62:12:55.88 &    A &    24.09 &     0.03 & 3-203.0         \\
  78 &    12:36:50.84 &    62:12:51.54 &    A &    25.20 &     0.05 & 3-259.0         \\
  79 &    12:36:50.84 &    62:12:27.24 &    A &    27.44 &     0.18 & 3-528.0         \\
  80 &    12:36:51.03 &    62:12:54.75 &    A &    27.56 &     0.15 & 3-208.0         \\
  81 &    12:36:51.06 &    62:13:20.60 &    A &    21.71 &     0.02 & 2-404.0         \\
  82 &    12:36:51.44 &    62:13: 0.26 &    A &    24.16 &     0.04 & 3-174.0         \\
  83 &    12:36:51.71 &    62:12:20.25 &    A &    25.92 &     0.09 & 3-659.0         \\
  84 &    12:36:51.76 &    62:13:53.81 &    A &    26.13 &     0.14 & 2-652.0         \\
  85\tablenotemark{d} &    12:36:51.95 &    62:11:55.53 &    A &    25.47 &     0.08 & 3-956.0         \\
  86 &    12:36:51.96 &    62:12:30.52 &    A &    28.75 &     0.20 & 3-523.0         \\
  87 &    12:36:52.02 &    62:12: 9.72 &    A &    24.78 &     0.04 & 3-777.0         \\
  88 &    12:36:52.03 &    62:14: 0.96 &    A &    26.58 &     0.10 & 2-702.0         \\
  89 &    12:36:52.21 &    62:13:23.34 &    A &    27.74 &     0.24 & 2-486.0         \\
  90 &    12:36:52.23 &    62:13:48.06 &    A &    27.03 &     0.12 & 2-646.0         \\
  91 &    12:36:52.36 &    62:13:46.68 &    A &    28.71 &     0.25 & 2-640.0         \\
  92 &    12:36:52.69 &    62:12:19.69 &    A &    25.33 &     0.05 & 3-696.0         \\
  93\tablenotemark{f} &    12:36:52.78 &    62:13:56.07 &    A &    27.67 &     0.14 & 2-736.2         \\
  94 &    12:36:52.91 &    62:14: 8.51 &    A &    26.96 &     0.10 & 2-834.0         \\
  95 &    12:36:52.98 &    62:12:56.76 &    A &    26.70 &     0.09 & 3-271.0         \\
  96 &    12:36:53.11 &    62:12:56.95 &    A &    26.36 &     0.13 & 3-271.1         \\
  97 &    12:36:53.23 &    62:13:43.60 &    A &    27.89 &     0.21 & 2-712.0         \\
  98 &    12:36:53.33 &    62:13: 0.59 &    A &    27.54 &     0.15 & 3-227.0         \\
  99\tablenotemark{d} &    12:36:53.39 &    62:13:25.05 &    A &    27.99 &     0.20 & 2-619.0         \\
  100 &    12:36:53.46 &    62:12:34.23 &    A &    26.38 &     0.09 & 3-551.0         \\
  101 &    12:36:53.48 &    62:12:20.61 &    A &    26.48 &     0.12 & 3-708.0         \\
  102 &    12:36:53.49 &    62:12:10.93 &    A &    27.81 &     0.27 & 3-801.0         \\
  103\tablenotemark{g} &    12:36:54.03 &    62:12:45.70 &    A &    27.07 &     0.12 & 3-419.0         \\
  104 &    12:36:54.71 &    62:13: 9.35 &    A &    28.57 &     0.21 & 3-170.0         \\
  105 &    12:36:54.73 &    62:13:30.33 &    A &    27.14 &     0.20 & 2-802.112       \\
  106 &    12:36:54.79 &    62:12:58.19 &    A &    27.71 &     0.21 & 3-318.0         \\
  107 &    12:36:55.01 &    62:13:14.75 &    A &    25.92 &     0.07 & 3-132.0         \\
  108 &    12:36:55.07 &    62:13:29.13 &    A &    27.92 &     0.07 & 2-802.1112      \\
  109 &    12:36:55.14 &    62:13:11.36 &    A &    24.81 &     0.03 & 3-180.2         \\
  110 &    12:36:55.25 &    62:12:52.43 &    A &    26.07 &     0.10 & 3-398.0         \\
  111 &    12:36:55.27 &    62:13: 9.50 &    A &    28.18 &     0.19 & 3-187.0         \\
  112 &    12:36:55.42 &    62:12:27.95 &    A &    28.36 &     0.25 & 3-695.0         \\
  113 &    12:36:55.59 &    62:13:59.89 &    A &    26.34 &     0.11 & 2-1018.0        \\
  114 &    12:36:55.78 &    62:13:48.78 &    A &    28.11 &     0.26 & 2-966.0         \\
  115 &    12:36:56.11 &    62:12:41.25 &    A &    28.04 &     0.18 & 3-610.111112    \\
  116 &    12:36:56.41 &    62:12: 9.22 &    A &    25.24 &     0.05 & 3-943.0         \\
  117 &    12:36:56.63 &    62:12:44.71 &    A &    25.32 &     0.19 & 3-610.1111111   \\
  118 &    12:36:56.95 &    62:12:58.24 &    A &    26.90 &     0.18 & 3-404.0         \\
  119 &    12:36:57.23 &    62:12:25.87 &    A &    24.42 &     0.04 & 3-773.0         \\
  120 &    12:36:57.32 &    62:12:59.71 &    A &    23.20 &     0.02 & 3-400.0         \\
  121 &    12:36:57.36 &    62:12:56.24 &    A &    27.16 &     0.17 & 3-412.0         \\
  122\tablenotemark{f} &    12:36:57.46 &    62:12:12.00 &    A &    25.11 &     0.04 & 3-965.111112    \\
  123 &    12:36:57.53 &    62:13:16.82 &    A &    26.91 &     0.10 & 3-184.0         \\
  124 &    12:36:58.00 &    62:12:25.04 &    A &    27.66 &     0.17 & 3-793.0         \\
  125 &    12:36:58.01 &    62:12:35.54 &    A &    27.32 &     0.22 & 3-698.0         \\
  126 &    12:36:58.07 &    62:13: 0.34 &    A &    23.73 &     0.02 & 3-405.0         \\
  127 &    12:36:58.17 &    62:13: 6.49 &    A &    26.01 &     0.07 & 3-342.0         \\
  128 &    12:36:58.31 &    62:12:51.09 &    A &    28.22 &     0.06 & 3-534.12        \\
  129 &    12:36:58.32 &    62:12:55.39 &    A &    27.82 &     0.16 & 3-454.12        \\
  130 &    12:36:58.36 &    62:12:56.34 &    A &    26.78 &     0.10 & 3-454.0         \\
  131 &    12:36:58.65 &    62:12:21.64 &    A &    26.42 &     0.10 & 3-863.0         \\
  132 &    12:36:58.70 &    62:12:17.04 &    A &    27.67 &     0.20 & 3-923.0         \\
  133\tablenotemark{d} &    12:36:58.76 &    62:12:52.46 &    A &    23.36 &     0.03 & 3-534.0         \\
  134 &    12:36:59.38 &    62:12:21.68 &    A &    25.75 &     0.08 & 3-908.1         \\
  135 &    12:36:59.53 &    62:12:21.11 &    A &    26.83 &     0.13 & 3-908.0         \\
  \enddata \tablenotetext{a}{Far-UV flux weighted position within HDF
    WFPC2 isophote} \tablenotetext{b}{A = ACS/SBC; S = STIS}
  \tablenotetext{c}{from catalog of \citep{Williams 1996}}
  \tablenotetext{d}{Lies on the edge of the far-UV image.}
  \tablenotetext{e}{Lies on the edge of the NICMOS images.}
  \tablenotetext{f}{NICMOS aperture includes multiple sources.}
  \tablenotetext{g}{WFPC2 aperture includes multiple sources.}
  \tablenotetext{h}{Star} \tablenotetext{i}{z=0.089 elliptical.  The
    FUV flux is limited to a much smaller aperture than the measured
    in the WFPC2 filters.  The F300W flux likely to be dominated by
    older stars.}
\end{deluxetable}

\clearpage

\begin{deluxetable}{lrrrr}
\tabletypesize{\scriptsize}
\tablecaption{SBC number counts\label{tbl: nc}}
\tablehead{
\colhead{$FUV_{AB}$} &
\colhead{n.c.} &
\colhead{low} &
\colhead{high} &
\colhead{Raw N} \\
\colhead{(mag.)} &
\colhead{N(deg$^{-2}$\ mag$^{-1}$)} &
\colhead{} &
\colhead{} &
\colhead{} 
}
\startdata

    19.5000 &     \nodata &    \nodata &    \nodata &     0 \\
    20.5000 &        1077 &        186 &       3553 &     1 \\
    21.5000 &        1089 &        188 &       3594 &     1 \\
    22.5000 &     \nodata &    \nodata &    \nodata &     0 \\
    23.5000 &        6202 &       3915 &       9543 &     6 \\
    24.5000 &        7890 &       5160 &      11787 &     8 \\
    25.5000 &       17103 &      12870 &      22533 &    17 \\
    26.5000 &       31595 &      25954 &      38357 &    31 \\
    27.5000 &       39479 &      32745 &      47485 &    34 \\
    28.5000 &       32524 &      23929 &      43745 &    14 \\

 \enddata

\end{deluxetable}

\clearpage

\clearpage

\clearpage

\begin{figure}[t*]
\caption{ \label{fig: FUV-mosaic} A color composite of the FUV and F450W images of the HDF.  The 
background image is the WFPC2 F450W image, over which the FUV data from STIS and ACS/SBC are shown 
in magenta.}
\end{figure}

\clearpage

\clearpage

\clearpage

\clearpage

\clearpage

\clearpage

\clearpage

\begin{figure}[t*]
\caption{ \label{fig: morph kcorr} The morphological K-correction.
We compare the UV morphology of galaxies at different redshifts to their
appearance in WFPC2 F300W and F814W bands, as well as the NICMOS F160W. The 
FUV and F300W data have been smoothed with a Gaussian kernel corresponding
to the FWHM of a point source.}
\end{figure}

\clearpage

\clearpage

\clearpage

\clearpage

\clearpage

\end{document}